\documentclass[draftclsnofoot, onecolumn]{IEEEtran}

\usepackage{vwcol}
\usepackage{url}
\usepackage{graphicx}
\usepackage{psfrag}
\usepackage{amsmath}
\usepackage{amsthm}
\usepackage{amsfonts}
\usepackage{amssymb}
\usepackage{bm}

\usepackage[ruled,linesnumbered]{algorithm2e}

\usepackage{algpseudocode}
\usepackage{xcolor}
\usepackage{float}
\usepackage{bbm}
\usepackage{mathrsfs}
\usepackage{ulem}
\usepackage{bbold}
\usepackage[justification=centering]{caption}
\usepackage{subfigure}
\usepackage{cite}
\usepackage{hyperref} 
\usepackage{multicol}
\usepackage{tikz}
\usetikzlibrary{arrows.meta}
\usepackage{setspace}
\usepackage{color}
\usepackage[numbers,sort&compress]{natbib}
\usepackage{latexsym}
\usepackage{extarrows}
\usepackage{cancel}
\usepackage{lipsum}

\usepackage{pifont}
\usepackage{multirow}
\usepackage{indentfirst}
\usepackage{color}
\usepackage{ulem}

\usepackage{cuted}

\usepackage{tcolorbox}

\usepackage{makecell}

\usepackage{colortbl}
\usepackage{booktabs}
\usepackage{arydshln}

\usepackage{textcomp}
\usepackage{manyfoot}
\usepackage{listings}
\usepackage{pifont}

\newcommand{\bs}{\boldsymbol}
\newcommand{\Sf}{\mathsf}
\newcommand{\TB}{\textbf}

\newtheorem{assumption}{Assumption}
\newtheorem{lemma}{Lemma}

\newtheorem{proposition}{Proposition}

\newcommand{\Eye}{\bs{I}}

\newcommand{\Trace}{\Sf{Tr}}
\newcommand{\Ts}{\Sf{T}}
\newcommand{\Cts}{\Sf{H}}

\newcommand{\Normal}{\Sf{N}}

\newcommand{\dd}{\Sf{d}}

\newcommand{\Proj}{\Sf{Proj}}
\newcommand{\Diag}{\Sf{D}}
\newcommand{\diag}{\Sf{d}}

\newcommand{\Mean}{\mathbb{E}}
\newcommand{\Var}{\mathbb{V}}

\newcommand{\Real}{\mathbb{R}}

\newcommand{\EXP}{\Sf{e}}

\newcommand{\Extr}{\mathop{\operatorname{extr}}}

\newcommand{\ArgMin}{\mathop{\operatorname{argmin}}}
\newcommand{\ArgMax}{\mathop{\operatorname{argmax}}}

\newcounter{magicrownumbers}
\newcommand\rownumber{\stepcounter{magicrownumbers}\arabic{magicrownumbers}}

\allowdisplaybreaks[4]

\ifCLASSINFOpdf

\else

\fi

\begin{document}

\title{
	K-step Vector Approximate Survey Propagation
}

\author{
	Qun Chen, Haochuan Zhang*, and Huimin Zhu
	\thanks{
		Q. Chen and H. Zhang are with Guangdong University of Technology, Guangzhou 510006, China (Emails:
		c.prey.q@gmail.com;
		haochuan.zhang@gdut.edu.cn).
		H. Zhu is with Guangzhou University of Chinese Medicine, Guangzhou 510006, China (email:
		hm\_zhu@gzucm.edu.cn).
		This work was supported by Guangdong Basic and Applied Basic Research Foundation under Grants 2022A1515010196 and 2023A1515110853.
	}
	\thanks{
		*Corresponding author:
		Haochuan Zhang.
	}
}

\maketitle

\begin{abstract}
Approximate Message Passing (AMP), originally developed to address high-dimensional linear inverse problems, has found widespread applications in signal processing and statistical inference.
Among its notable variants, Vector Approximate Message Passing (VAMP), Generalized Approximate Survey Propagation (GASP), and Vector Approximate Survey Propagation (VASP) have demonstrated effectiveness even when the assumed generative models differ from the true models.
However, many fundamental questions regarding model mismatch remain unanswered.
For instance, it is still unclear what level of model mismatch is required for the postulated posterior estimate (PPE) to exhibit a replica symmetry breaking (RSB) structure in the extremum conditions of its free energy, and what order of RSB is necessary.
In this paper, we introduce a novel approximate message passing algorithm that incorporates K-step RSB (KRSB) and naturally reduces to VAMP and VASP with specific parameter selections.
We refer to this as the K-step VASP (KVASP) algorithm.
Simulations show that KVASP significantly outperforms VAMP and GASP in estimation accuracy, particularly when the assumed prior has discrete support and the measurement matrix is non-i.i.d..
Additionally, the state evolution (SE) of KVASP, derived heuristically, accurately tracks the per-iteration mean squared error (MSE).
A comparison between the SE and the free energy under the KRSB ansatz reveals that the fixed-point equations of SE align with the saddle-point equations of the free energy.
This suggests that, once the KRSB ansatz holds and the SE fixed point is reached, KVASP can accurately compute the PPE in the large system limit (LSL).
\end{abstract}


\IEEEpeerreviewmaketitle

\section{Introduction}
In this paper, we focus on estimating a high-dimensional signal derived from observations generated by a generalized linear model (GLM) \cite{takahashi2022macroscopic,lucibello2019generalized,chen2023vector}.
Specifically, we consider an
$ N \times 1 $
random signal
$
\bs{x}_{0}
\sim
p(\bs{x}_{0})
$,
which is linearly transformed into
$
\bs{z}_{0}
\triangleq
\bs{H} \bs{x}_{0}
$
using a
deterministic measurement matrix
$ \bs{H} $
of size
$ M \times N $.
The resulting
$ M \times 1 $
vector is then randomly mapped to yield the observation
$
\bs{y}
\sim
p(\bs{y} | \bs{z}_{0})
$,
i.e.,
\begin{align*}
p( \bs{x}_{0} )
& \triangleq
\prod_{i = 1}^{N}
p( x_{0, i} )
, \\
p( \bs{y} | \bs{z}_{0} )
& \triangleq
\prod_{a = 1}^{M}
p( y_{a} | z_{0, a} )
.
\end{align*}
Following the approach in \cite{takahashi2022macroscopic,lucibello2019generalized,chen2023vector}, we examine a model-mismatched setting for estimating the input signal
$ \bs{x}_{0} $.
In this scenario, we assume that the prior and likelihood information utilized by the estimator differ from the true ones in the GLM.
We denote the assumed prior and likelihood as
$ q( \bs{x}_{0} ) $
and
$ q( \bs{y} | \bs{z}_{0} ) $,
where
\begin{align*}
q( \bs{x}_{0} )
& \triangleq
\prod_{i = 1}^{N}
q( x_{0, i} )
, \\
q( \bs{y} | \bs{z}_{0} )
& \triangleq
\prod_{a = 1}^{M}
q( y_{a} | z_{0, a} )
.
\end{align*}
We can represent the postulated posterior estima (PPE) as follows \cite{takahashi2022macroscopic,lucibello2019generalized,antenucci2019approximate,chen2023vector,rangan2012asymptotic}:
\begin{align}
\hat{\bs{x}}
& =
\int{
	\dd \bs{z}_{0}
	\dd \bs{x}_{0}
} \,
q( \bs{z}_{0}, \bs{x}_{0} )
\bs{x}_{0}
, \label{Eq:Postulated_Posterior_Estimator} \\
q( \bs{z}_{0}, \bs{x}_{0} )
& =
\frac{1}{ Z(\bs{y}) }
q^{\beta}( \bs{y} | \bs{z}_{0} )
\delta(
	\bs{z}_{0}
	-
	\bs{H} \bs{x}_{0}
)
q^{\beta}(\bs{x}_{0})
, \label{Eq:Postulated_Posterior} \\
Z( \bs{y} )
& =
\int{
	\dd \bs{z}_{0}
	\dd \bs{x}_{0}
} \,
q^{\beta}( \bs{y} | \bs{z}_{0} )
\delta(
	\bs{z}_{0}
	-
	\bs{H} \bs{x}_{0}
)
q^{\beta}( \bs{x}_{0} )
. \label{Eq:Normalization_Factor}
\end{align}
Here, the estimate
$ \hat{\bs{x}} $
is referred to as the postulated MAP (PMAP) estimate \cite{rangan2012asymptotic} when
$ \beta \to + \infty $,
and as the postulated minimum mean squared error (PMMSE) estimate \cite{rangan2012asymptotic} when
$ \beta = 1 $.
In the large system limit (LSL), where
$ M \rightarrow \infty $
and
$ N \rightarrow \infty $
with a fixed ratio
$ \alpha = \frac{M}{N} $,
solving the PMAP estimation problem becomes highly non-trivial.
In fact, this estimation is non-convex and NP-hard in the worst case \cite{antenucci2019approximate}.

The challenge of exact computation has led to the development of numerous approximate algorithms, one of which is approximate message passing (AMP).
AMP refers to a family of iterative algorithms that have been widely applied in contexts such as linear regression \cite{donoho2009message,krzakala2012probabilistic,kabashima2003cdma,ma2017orthogonal,rangan2019vector,opper2005expectation,minka2013expectation,liu2021memory}, generalized linear regression \cite{rangan2011generalized,schniter2016vector,he2017generalized,tian2021generalized}, and low-rank matrix factorization \cite{kabashima2016phase,montanari2021estimation,barbier2023fundamental}.
The popularity of AMP can be attributed to several key features \cite{depope2023light}:
AMP leverages the structural information of the signal;
under mild model assumptions, its performance is accurately described by a set of scalar recursive equations known as state evolution (SE) \cite{bayati2011dynamics,javanmard2013state};
using SE, it can be shown that AMP achieves Bayes-optimal performance in various settings \cite{montanari2021estimation,barbier2023fundamental,barbier2019optimal}.
However, these advantages of AMP rely on a critical assumption that the estimator has perfect knowledge of the prior and likelihood, which may not be feasible in certain situations.

A more practical task is to study estimation in the model-mismatched setting, where the distributions provided to the estimator may be incorrect.
Unfortunately, only a limited number of results are currently available in this area.
In \cite{antenucci2019approximate}, Antenucci \textit{et al.} demonstrated that the performance of Low-RAMP \cite{lesieur2017constrained} degrades rapidly in the presence of model mismatch.
To address this issue, they developed a new algorithm called approximate survey propagation (ASP) for the matrix factorization task, which accounts for the glassy nature of the underlying problem.
Compared to Low-RAMP, the ASP algorithm converges over a larger regime and achieves significantly lower mean squared error (MSE).
Notably, its saddle point equations reproduce the one-step replica symmetry breaking (1RSB) equations of the system's free energy, a crucial quantity in the physics of disordered systems.
In \cite{lucibello2019generalized}, Lucibello \textit{et al.} also showed that model mismatch significantly degrades the performance of the generalized AMP (GAMP) \cite{rangan2011generalized}.
They proposed the generalized ASP (GASP) algorithm to approximate the PPE in \eqref{Eq:Postulated_Posterior_Estimator}, which was later found to generally outperform GAMP.
The saddle point equations derived for GASP effectively capture the dynamics of the algorithm in the LSL.
However, the derivation of GASP requires that the elements of the measurement matrix
$\bs{H}$
be independent and identically distributed (i.i.d.), leaving the situation for non-i.i.d. cases unclear.
To address the limitations of GASP, \cite{chen2023vector} introduced a new algorithm called Vector ASP (VASP).
VASP employs vector-form surveys to effectively capture correlations within the measurement matrix.
In VASP, the fixed point equations of this SE are identical to the saddle point equations of the free energy, as established by \cite{takahashi2022macroscopic} under the 1RSB ansatz.
This connection between the SE and the 1RSB analysis implies that if the 1RSB ansatz holds true and the algorithm converges to the SE fixed point, VASP efficiently implements the PPE in \eqref{Eq:Postulated_Posterior_Estimator} with only cubic computational complexity.

However, in the context of statistical inference under model mismatch, algorithms such as VAMP \cite{takahashi2022macroscopic}, GASP\cite{lucibello2019generalized}, and VAS \cite{chen2023vector} can provide accurate estimates in certain scenarios.
Nonetheless, many fundamental questions about model mismatch remain unresolved.
For example, it is still unclear what degree of model mismatch is necessary for the PPE in \eqref{Eq:Postulated_Posterior_Estimator} to exhibit a replica symmetry breaking (RSB) structure in the extremum conditions of its free energy \cite{takahashi2022macroscopic}, and what order of RSB is adequate.
In this paper, our primary contribution is the development of a novel approximate message passing algorithm that incorporates K-step RSB (KRSB) and seamlessly reduces to VAMP and VASP under specific parameter selections.
We refer to this as the K-step VASP (KVASP) algorithm, building on the impactful research in survey propagation related to sparse random constraint satisfaction problems \cite{mezard2002analytic,braunstein2005survey}.
The dynamics of KVASP are well characterized by the SE we present.
Interestingly, the SE fixed point equations align with the saddle points of the KRSB free energy.
This correspondence suggests that, once the KRSB ansatz is valid and the SE fixed point is reached, KVASP can accurately compute the PPE in \eqref{Eq:Postulated_Posterior_Estimator} in LSL.

Throughout this paper, we use the following notation conventions:
Plain letters (e.g.,
$ m $
and
$ v $)
represent scalars.
Boldface lowercase letters (e.g.,
$ \bs{m} $
and
$ \bs{v} $)
represent column vectors.
Boldface lowercase letters with a vector arrow (e.g.,
$ \vec{\bs{m}} $
and
$ \vec{\bs{v}} $)
represent row vectors.
Boldface uppercase letters (e.g.,
$ \bs{V} $
and
$ \bs{C} $)
represent matrices.
$ \EXP(\cdot) $
denotes the exponential operator.
$ \delta(\cdot) $
represents a Dirac delta function.
$ \Normal[ \bs{x} | \bs{m}, \bs{C} ] $
denotes a Gaussian distribution with mean
$ \bs{m} $
and covariance
$ \bs{C} $,
defined as
$
\Normal[ \bs{x} | \bs{m}, \bs{C} ]
\triangleq
| 2 \pi \bs{C} |^{ - \frac{1}{2} }
\EXP[
	-
	\frac{1}{2}
	( \bs{x} - \bs{m} )^{\Ts}
	\bs{C}^{-1}
	( \bs{x} - \bs{m} )
]
$.
For any matrix
$ \bs{A} $,
$ A^{ ( i, j ) } $
represents the element at the
$ i $-th row and
$ j $-th column,
$ \bs{A}^{\Ts} $
denotes the transpose of matrix
$ \bs{A} $,
$ \Trace(\bs{A}) $
represents the trace of matrix
$ \bs{A} $.
$ \bs{e}_{N} $
is a
$ N $-dimensional column vector with all zeros except for a
$ 1 $
in the first element.
$ \bs{1}_{N} $
is a
$ N $-dimensional column vector consisting of all ones.
$\Eye_{N}$
represents the identity matrix of size
$ N $.
$ \mathbb{1}_{N} $
is a matrix of size
$ N $
with all entries equal to one.
$ \Diag(\bs{v}) $
is a diagonal matrix with diagonal elements equal to the entries of vector
$ \bs{v} $.
$ \diag(\bs{C}) $
is a diagonal operator that returns an
$ N $-dimensional column vector containing the diagonal elements of the matrix
$ \bs{C} $.

\section{Derivation of this KRSB Free Energy}

In the section, we outline the KRSB calculation of the free energy.
Given that analogous calculations can be found in \cite{takahashi2022macroscopic,shinzato2008perceptron,shinzato2008learning,kabashima2008inference}, we only describe the main steps.
The average free energy of the inference problem \eqref{Eq:Postulated_Posterior_Estimator} is defined as \cite{takahashi2022macroscopic}:
\begin{align*}
f
& \triangleq
-
\lim_{ N \rightarrow \infty }
\frac{1}{ N \beta }
\Mean_{
	\bs{y}, \bs{H}, \bs{x}_{0}
}[
	\log Z( \bs{y} )
]
.
\end{align*}
This average free energy serves as the cumulant generating function \cite{mezard2009information,lewis1989large}, capturing all the cumulants of the Boltzmann distribution \eqref{Eq:Postulated_Posterior}.
Using the replica method, the average free energy is expressed as follows \cite{takahashi2022macroscopic}:
\begin{align*}
f
& =
-
\lim_{ \tau \rightarrow 0 }
\frac{1}{\tau}
\Extr_{
	\bs{Q}_{\Sf{z}},
	\bs{Q}_{\Sf{x}}
}[
	g_{\Sf{z}}
	+
	g_{\Sf{h}}
	+
	g_{\Sf{x}}
	-
	g_{\Sf{e}}
]
, \\
g_{\Sf{z}}
& \triangleq
\Extr_{
	\tilde{\bs{Q}}_{\Sf{z}}
}[
	\frac{\alpha}{\beta}
	\log \int{
		\dd y
		\dd \vec{\bs{z}}
	} \,
	p(
		y \vec{\bs{1}} | \vec{\bs{z}}
	)
	\EXP(
		-
		\frac{1}{2}
		\vec{\bs{z}}
		\tilde{\bs{Q}}_{\Sf{z}}
		\vec{\bs{z}}^{\Ts}
	)
	+
	\frac{\alpha}{2 \beta}
	\Trace(
		\bs{Q}_{\Sf{z}}
		\tilde{\bs{Q}}_{\Sf{z}}
	)
]
, \\
g_{\Sf{h}}
& \triangleq
\Extr_{
	\bs{\Lambda}_{\Sf{z}},
	\bs{\Lambda}_{\Sf{x}}
}[
	\frac{\alpha}{2 \beta}
	\Trace(
		\bs{Q}_{\Sf{z}}
		\bs{\Lambda}_{\Sf{z}}
	)
	-
	\frac{1}{2 \beta}
	\Mean_{\lambda}[
		\log
		|
			\bs{\Lambda}_{\Sf{x}}
			+
			\lambda \bs{\Lambda}_{\Sf{z}}
		|
	]	
	+
	\frac{1}{2 \beta}
	\Trace(
		\bs{Q}_{\Sf{x}}
		\bs{\Lambda}_{\Sf{x}}
	)
]
, \\
g_{\Sf{x}}
& \triangleq
\Extr_{
	\tilde{\bs{Q}}_{\Sf{x}}
}[
	\frac{1}{\beta}
	\log
	\int{
		\dd \vec{\bs{x}}
	} \,
	\EXP(
		-
		\frac{1}{2}
		\vec{\bs{x}}
		\tilde{\bs{Q}}_{\Sf{x}}
		\vec{\bs{x}}^{\Ts}
	)
	p(
		\vec{\bs{x}}
	)
	+
	\frac{1}{2 \beta}
	\Trace(
		\bs{Q}_{\Sf{x}}
		\tilde{\bs{Q}}_{\Sf{x}}
	)
]
, \\
g_{\Sf{e}}
& \triangleq
\frac{\alpha}{2 \beta}
\log
| \bs{Q}_{\Sf{z}} |
+
\frac{1}{2 \beta}
\log
| \bs{Q}_{\Sf{x}} |
.
\end{align*}
Here,
$ \bs{Q}_{\Sf{z}} $,
$ \bs{Q}_{\Sf{x}} $,
$ \bs{\Lambda}_{\Sf{z}} $,
$ \bs{\Lambda}_{\Sf{x}} $,
$ \tilde{\bs{Q}}_{\Sf{z}} $,
$ \tilde{\bs{Q}}_{\Sf{x}} $
are
$ (\tau + 1) \times (\tau + 1) $
real symmetric matrices,
$ \Extr[ \cdot ] $
denotes an extremum operator,
$ \lambda $
follows the limiting eigenvalue distribution of
$ \bs{H}^{\Ts} \bs{H} $,
and the integration measures
$ \dd \vec{\bs{z}} $
and
$ \dd \vec{\bs{x}} $
are defined as:
$
\dd \vec{\bs{z}}
\triangleq
\dd z_{0}
\dd \tilde{z}_{1}
\cdots
\dd \tilde{z}_{\tau}
$,
$
\dd \vec{\bs{x}}
\triangleq
\dd x_{0}
\dd \tilde{x}_{1}
\cdots
\dd \tilde{x}_{\tau}
$
with
$
\vec{\bs{z}}
\triangleq
[
	z_{0},
	\tilde{z}_{1},
	\cdots,
	\tilde{z}_{\tau}
]
$,
$
\vec{\bs{x}}
\triangleq
[
	x_{0},
	\tilde{x}_{1},
	\cdots,
	\tilde{x}_{\tau}
]
$,
$
p(
	y \vec{\bs{1}}_{\tau + 1} |
	\vec{\bs{z}}
)
\triangleq
p( y | z_{0} )
\prod_{i = 1}^{\tau}
q^{\beta}( y | \tilde{z}_{i} )
$,
and
$
p( \vec{\bs{x}} )
\triangleq
p( x_{0} )
\prod_{i = 1}^{\tau}
q^{\beta}( \tilde{x}_{i} )
$.
The extremum conditions are then formulated as \cite{takahashi2022macroscopic}:
\begin{subequations}
\begin{align}
\bs{0}
& =
\bs{\Lambda}_{\Sf{z}}
-
\bs{Q}_{\Sf{z}}^{-1}
+
\tilde{\bs{Q}}_{\Sf{z}}
, \label{Eq:Extr_Cond:Z} \\
\bs{0}
& =
\bs{\Lambda}_{\Sf{x}}
-
\bs{Q}_{\Sf{x}}^{-1}
+
\tilde{\bs{Q}}_{\Sf{x}}
, \label{Eq:Extr_Cond:X} \\
\bs{Q}_{\Sf{z}}
& =
\frac{1}{\alpha}
\Mean_{\lambda}[
	\lambda (
		\bs{\Lambda}_{\Sf{x}}
		+
		\lambda
		\bs{\Lambda}_{\Sf{z}}
	)^{-1}
]
, \label{Eq:Extr_Cond:Exp_Z} \\
\bs{Q}_{\Sf{x}}
& =
\Mean_{\lambda}[
	(
		\bs{\Lambda}_{\Sf{x}}
		+
		\lambda
		\bs{\Lambda}_{\Sf{z}}
	)^{-1}
]
, \label{Eq:Extr_Cond:Exp_X} \\
\bs{Q}_{\Sf{z}}
& =
\frac{
	\int{
		\dd y
		\dd \vec{\bs{z}}
	} \,
	p(
		y \vec{\bs{1}} | \vec{\bs{z}}
	)
	\EXP(
		-
		\frac{1}{2}
		\vec{\bs{z}}
		\tilde{\bs{Q}}_{\Sf{z}}
		\vec{\bs{z}}^{\Ts}
	)
	\vec{\bs{z}}^{\Ts} \vec{\bs{z}}
}{
	\int{
		\dd y
		\dd \vec{\bs{z}}
	} \,
	p(
		y \vec{\bs{1}} | \vec{\bs{z}}
	)
	\EXP(
		-
		\frac{1}{2}
		\vec{\bs{z}}
		\tilde{\bs{Q}}_{\Sf{z}}
		\vec{\bs{z}}^{\Ts}
	)
}
, \label{Eq:Extr_Cond:Int_Z} \\
\bs{Q}_{\Sf{x}}
& =
\frac{
	\int{
		\dd \vec{\bs{x}}
	} \,
	\EXP(
		-
		\frac{1}{2}
		\vec{\bs{x}}
		\tilde{\bs{Q}}_{\Sf{x}}
		\vec{\bs{x}}^{\Ts}
	)
	p(
		\vec{\bs{x}}
	)
	\vec{\bs{x}}^{\Ts} \vec{\bs{x}}
}{
	\int{
		\dd \vec{\bs{x}}
	} \,
	\EXP(
		-
		\frac{1}{2}
		\vec{\bs{x}}
		\tilde{\bs{Q}}_{\Sf{x}}
		\vec{\bs{x}}^{\Ts}
	)
	p(
		\vec{\bs{x}}
	)
}
. \label{Eq:Extr_Cond:Int_X}
\end{align}
\label{Eq:Extreme_Condition}
\end{subequations}

Under the KRSB ansatz, the matrices of the saddle point equations follow a common structure, expressed as \cite{bereyhi2019statistical,abbara2020statistical,bereyhi2017replica}:
\begin{subequations}
\begin{align}
\bs{Q}_{\Sf{z}}
& =
\mathcal{Q}
(
	C_{\Sf{z}}, D_{\Sf{z}}, F_{\Sf{z}},
	\frac{1}{\beta} \chi_{\Sf{z}},
	\{
		H_{\Sf{z, k}}
	\}_{k = 1}^{K}
)
, \\
\bs{Q}_{\Sf{x}}
& =
\mathcal{Q}
(
	C_{\Sf{x}}, D_{\Sf{x}}, F_{\Sf{x}},
	\frac{1}{\beta} \chi_{\Sf{x}},
	\{
		H_{\Sf{x, k}}
	\}_{k = 1}^{K}
)
, \label{eq:def_Dx} \\
\bs{\Lambda}_{\Sf{z}}
& =
\mathcal{Q}
(
	\hat{C}_{\Sf{2z}},
	- \beta \hat{D}_{\Sf{2z}},
	- \beta^{2} \hat{F}_{\Sf{2z}},
	\beta \hat{\chi}_{\Sf{2z}},
	\{
		- \beta^{2} \hat{H}_{\Sf{2z, k}}
	\}_{k = 1}^{K}
)
, \\
\bs{\Lambda}_{\Sf{x}}
& =
\mathcal{Q}
(
	\hat{C}_{\Sf{2x}},
	- \beta \hat{D}_{\Sf{2x}},
	- \beta^{2} \hat{F}_{\Sf{2x}},
	\beta \hat{\chi}_{\Sf{2x}},
	\{
		- \beta^{2} \hat{H}_{\Sf{2x, k}}
	\}_{k = 1}^{K}
)
, \\
\tilde{\bs{Q}}_{\Sf{z}}
& =
\mathcal{Q}
(
	\hat{C}_{\Sf{1z}},
	- \beta \hat{D}_{\Sf{1z}},
	- \beta^{2} \hat{F}_{\Sf{1z}},
	\beta \hat{\chi}_{\Sf{1z}},
	\{
		- \beta^{2} \hat{H}_{\Sf{1z, k}}
	\}_{k = 1}^{K}
)
, \\
\tilde{\bs{Q}}_{\Sf{x}}
& =
\mathcal{Q}
(
	\hat{C}_{\Sf{1x}},
	- \beta \hat{D}_{\Sf{1x}},
	- \beta^{2} \hat{F}_{\Sf{1x}},
	\beta \hat{\chi}_{\Sf{1x}},
	\{
		- \beta^{2} \hat{H}_{\Sf{1x, k}}
	\}_{k = 1}^{K}
)
,
\end{align}
\end{subequations}
with
$
\mathcal{A}(
	F, \chi,
	\{
		H_{k}
	\}_{k = 1}^{K}
)
\triangleq
F \mathbb{1}_{\tau}
+
\chi \Eye_{\tau}
+
\sum_{k = 1}^{K}{}
H_{k} \Eye_{
	\frac{ \tau }{
		\tilde{L}_{k}
	}
}
\otimes
\mathbb{1}_{
	\tilde{L}_{k}
}
$,
$
\mathcal{Q}(
	C, D, F, \chi,
	\{
		H_{k}
	\}_{k = 1}^{K}
)
\triangleq
\left[
\begin{array}{cc}
	C
	&
	D \vec{\bs{1}}_{\tau}
	\\
	D \bs{1}_{\tau}
	&
	\mathcal{A}(
		F, \chi,
		\{
			H_{k}
		\}_{k = 1}^{K}
	)
	\\
\end{array}
\right]
$,
$
\tilde{L}_{k}
\triangleq
\frac{ L_{k} }{\beta}
$,
$
\{
	\tilde{L}_{k}
\}_{k = 1}^{K}
\in
\mathbb{Z}^{+}
$,
$
\{
	\frac{
		\tilde{L}_{k}
	}{
		\tilde{L}_{k - 1}
	}
\}_{k = 2}^{K}
\in
\mathbb{Z}^{+}
$.
We reformulate the extremum conditions by applying a Gaussian density convolution technique \cite{bromiley2003products}.
For convenience, all relevant definitions are summarized in Tab. \ref{Tab:notations} in Appendix \ref{Appendix:Summary}.
The KRSB saddle point equations are provided in Algo. \ref{Tab:Saddle_Point}.

\begin{algorithm}[!t]
\caption{The KRSB saddle point equations}
\label{Tab:Saddle_Point}
\scriptsize
\begin{algorithmic}[1]
\State
$
C_{\Sf{x}}
=
\Mean[ x_{0}^{2} ]
$
\\
$
\hat{C}_{\Sf{1x}}
=
0
$
\\
$
\hat{C}_{\Sf{2x}}
=
\frac{1}{ C_{\Sf{x}} }
$
\\
$
C_{\Sf{z}}
=
\frac{1}{\alpha}
\Mean_{\lambda}[
	\lambda
]
C_{\Sf{x}}
$
\\
$
\hat{C}_{\Sf{1z}}
=
\frac{1}{ C_{\Sf{z}} }
$
\\
$
\hat{C}_{\Sf{2z}}
=
0
$
\\
$
0
=
\hat{D}_{\Sf{1x}}
+
\hat{D}_{\Sf{2x}}
-
\frac{D_{\Sf{x}}}{
	C_{\Sf{x}}
	(
		\chi_{\Sf{x}}
		+
		\sum_{k = 1}^{K}{}
		L_{k} H_{\Sf{x, k}}
	)
}
$
\\
$
0
=
\hat{F}_{\Sf{1x}}
+
\hat{F}_{\Sf{2x}}
+
\frac{
	D_{\Sf{x}}^{2}
}{
	C_{\Sf{x}}
	(
		\chi_{\Sf{x}}
		+
		\sum_{k = 1}^{K}{}
		L_{k} H_{\Sf{x, k}}
	)^{2}
}
-
\frac{F_{\Sf{x}}}{
	(
		\chi_{\Sf{x}}
		+
		\sum_{k = 1}^{K}{}
		L_{k} H_{\Sf{x, k}}
	)^{2}
}
$
\\
$
0
=
\hat{\chi}_{\Sf{1x}}
+
\hat{\chi}_{\Sf{2x}}
-
\frac{1}{
	\chi_{\Sf{x}}
}
$
\\
$
0
=
\hat{H}_{\Sf{1x, 1}}
+
\hat{H}_{\Sf{2x, 1}}
+
\frac{1}{L_{1}}
(
	\frac{1}{
		\chi_{\Sf{x}}
		+
		L_{1} H_{\Sf{x, 1}}
	}
	-
	\frac{1}{
		\chi_{\Sf{x}}
	}
)
$
\\
$
0
=
\hat{H}_{\Sf{1x, k}}
+
\hat{H}_{\Sf{2x, k}}
+
\frac{1}{L_{k}}
(
	\frac{1}{
		\chi_{\Sf{x}}
		+
		\sum_{i = 1}^{k}{}
		L_{i} H_{\Sf{x, i}}
	}
	-
	\frac{1}{
		\chi_{\Sf{x}}
		+
		\sum_{i = 1}^{k - 1}{}
		L_{i} H_{\Sf{x, i}}
	}
)
, \,
k \in [ 2, K ]
$
\\
$
0
=
\hat{D}_{\Sf{1z}}
+
\hat{D}_{\Sf{2z}}
-
\frac{D_{\Sf{z}}}{
	C_{\Sf{z}}
	(
		\chi_{\Sf{z}}
		+
		\sum_{k = 1}^{K}{}
		L_{k} H_{\Sf{z, k}}
	)
}
$
\\
$
0
=
\hat{F}_{\Sf{1z}}
+
\hat{F}_{\Sf{2z}}
+
\frac{
	D_{\Sf{z}}^{2}
}{
	C_{\Sf{z}}
	(
		\chi_{\Sf{z}}
		+
		\sum_{k = 1}^{K}{}
		L_{k} H_{\Sf{z, k}}
	)^{2}
}
-
\frac{F_{\Sf{z}}}{
	(
		\chi_{\Sf{z}}
		+
		\sum_{k = 1}^{K}{}
		L_{k} H_{\Sf{z, k}}
	)^{2}
}
$
\\
$
0
=
\hat{\chi}_{\Sf{1z}}
+
\hat{\chi}_{\Sf{2z}}
-
\frac{1}{
	\chi_{\Sf{z}}
}
$
\\
$
0
=
\hat{H}_{\Sf{1z, 1}}
+
\hat{H}_{\Sf{2z, 1}}
+
\frac{1}{L_{1}}
(
	\frac{1}{
		\chi_{\Sf{z}}
		+
		L_{1} H_{\Sf{z, 1}}
	}
	-
	\frac{1}{
		\chi_{\Sf{z}}
	}
)
$
\\
$
0
=
\hat{H}_{\Sf{1z, k}}
+
\hat{H}_{\Sf{2z, k}}
+
\frac{1}{L_{k}}
(
	\frac{1}{
		\chi_{\Sf{z}}
		+
		\sum_{i = 1}^{k}{}
		L_{i} H_{\Sf{z, i}}
	}
	-
	\frac{1}{
		\chi_{\Sf{z}}
		+
		\sum_{i = 1}^{k - 1}{}
		L_{i} H_{\Sf{z, i}}
	}
)
, \,
k \in [ 2, K ]
$
\\
$
D_{\Sf{x}}
=
C_{\Sf{x}}
\Mean_{\lambda}[
	\frac{
		\hat{D}_{\Sf{2x}}
		+
		\lambda \hat{D}_{\Sf{2z}}
	}{
		(
			\hat{\chi}_{\Sf{2x}}
			-
			\sum_{k = 1}^{K}{}
			L_{k} \hat{H}_{\Sf{2x, k}}
		)
		+
		\lambda
		(
			\hat{\chi}_{\Sf{2z}}
			-
			\sum_{k = 1}^{K}{}
			L_{k} \hat{H}_{\Sf{2z, k}}
		)
	}
]
$
\\
$
F_{\Sf{x}}
=
C_{\Sf{x}}
\Mean_{\lambda}[
	\frac{
		(
			\hat{D}_{\Sf{2x}}
			+
			\lambda\hat{D}_{\Sf{2z}}
		)^{2}
	}{
		[
			(
				\hat{\chi}_{\Sf{2x}}
				-
				\sum_{k = 1}^{K}{}
				L_{k} \hat{H}_{\Sf{2x, k}}
			)
			+
			\lambda
			(
				\hat{\chi}_{\Sf{2z}}
				-
				\sum_{k = 1}^{K}{}
				L_{k} \hat{H}_{\Sf{2z, k}}
			)
		]^{2}
	}
]
+
\Mean_{\lambda}[
	\frac{
		\hat{F}_{\Sf{2x}}
		+
		\lambda \hat{F}_{\Sf{2z}}
	}{
		[
			(
				\hat{\chi}_{\Sf{2x}}
				-
				\sum_{k = 1}^{K}{}
				L_{k} \hat{H}_{\Sf{2x, k}}
			)
			+
			\lambda
			(
				\hat{\chi}_{\Sf{2z}}
				-
				\sum_{k = 1}^{K}{}
				L_{k} \hat{H}_{\Sf{2z, k}}
			)		
		]^{2}
	}
]
$
\\
$
\chi_{\Sf{x}}
=
\Mean_{\lambda}[
	\frac{1}{
		\hat{\chi}_{\Sf{2x}}
		+
		\lambda
		\hat{\chi}_{\Sf{2z}}
	}
]
$
\\
$
H_{\Sf{x, 1}}
=
\frac{1}{ L_{1} }
(
	\Mean_{\lambda}[
		\frac{1}{
			(
				\hat{\chi}_{\Sf{2x}}
				-
				L_{1} \hat{H}_{\Sf{2x, 1}}
			)
			+
			\lambda
			(
				\hat{\chi}_{\Sf{2z}}
				-
				L_{1} \hat{H}_{\Sf{2z, 1}}
			)	
		}
	]
	-
	\Mean_{\lambda}[
		\frac{1}{
			\hat{\chi}_{\Sf{2x}}
			+
			\lambda
			\hat{\chi}_{\Sf{2z}}
		}
	]
)
$
\\
$
H_{\Sf{x, k}}
=
\frac{1}{ L_{k} }
(
	\Mean_{\lambda}[
		\frac{1}{
			(
				\hat{\chi}_{\Sf{2x}}
				-
				\sum_{i = 1}^{k}{}
				L_{i} \hat{H}_{\Sf{2x, i}}
			)
			+
			\lambda
			(
				\hat{\chi}_{\Sf{2z}}
				-
				\sum_{i = 1}^{k}{}
				L_{i} \hat{H}_{\Sf{2z, i}}
			)	
		}
	]
	-
	\Mean_{\lambda}[
		\frac{1}{
			(
				\hat{\chi}_{\Sf{2x}}
				-
				\sum_{i = 1}^{k - 1}{}
				L_{i} \hat{H}_{\Sf{2x, i}}
			)
			+
			\lambda
			(
				\hat{\chi}_{\Sf{2z}}
				-
				\sum_{i = 1}^{k - 1}{}
				L_{i} \hat{H}_{\Sf{2z, i}}
			)	
		}
	]
)
, \,
k \in [ 2, K ]
$
\\
$
D_{\Sf{z}}
=
\frac{1}{\alpha}
C_{\Sf{x}}
\Mean_{\lambda}[
	\frac{
		\lambda
		(
			\hat{D}_{\Sf{2x}}
			+
			\lambda \hat{D}_{\Sf{2z}}
		)
	}{
		(
			\hat{\chi}_{\Sf{2x}}
			-
			\sum_{k = 1}^{K}{}
			L_{k} \hat{H}_{\Sf{2x, k}}
		)
		+
		\lambda
		(
			\hat{\chi}_{\Sf{2z}}
			-
			\sum_{k = 1}^{K}{}
			L_{k} \hat{H}_{\Sf{2z, k}}
		)
	}
]
$
\\
$
F_{\Sf{z}}
=
\frac{1}{\alpha}
C_{\Sf{x}}
\Mean_{\lambda}[
	\frac{
		\lambda
		(
			\hat{D}_{\Sf{2x}}
			+
			\lambda \hat{D}_{\Sf{2z}}
		)^{2}
	}{
		[
			(
				\hat{\chi}_{\Sf{2x}}
				-
				\sum_{k = 1}^{K}{}
				L_{k} \hat{H}_{\Sf{2x, k}}
			)
			+
			\lambda
			(
				\hat{\chi}_{\Sf{2z}}
				-
				\sum_{k = 1}^{K}{}
				L_{k} \hat{H}_{\Sf{2z, k}}
			)		
		]^{2}
	}
]
+
\frac{1}{\alpha}
\Mean_{\lambda}[
	\frac{
		\lambda
		(
			\hat{F}_{\Sf{2x}}
			+
			\lambda \hat{F}_{\Sf{2z}}
		)
	}{
		[
			(
				\hat{\chi}_{\Sf{2x}}
				-
				\sum_{k = 1}^{K}{}
				L_{k} \hat{H}_{\Sf{2x, k}}
			)
			+
			\lambda
			(
				\hat{\chi}_{\Sf{2z}}
				-
				\sum_{k = 1}^{K}{}
				L_{k} \hat{H}_{\Sf{2z, k}}
			)		
		]^{2}
	}
]
$
\\
$
\chi_{\Sf{z}}
=
\frac{1}{\alpha}
\Mean_{\lambda}[
	\frac{\lambda}{
		\hat{\chi}_{\Sf{2x}}
		+
		\lambda
		\hat{\chi}_{\Sf{2z}}
	}
]
$
\\
$
H_{\Sf{z, 1}}
=
\frac{1}{ \alpha L_{1} }
(
	\Mean_{\lambda}[
		\frac{\lambda}{
			(
				\hat{\chi}_{\Sf{2x}}
				-
				L_{1} \hat{H}_{\Sf{2x, 1}}
			)
			+
			\lambda
			(
				\hat{\chi}_{\Sf{2z}}
				-
				L_{1} \hat{H}_{\Sf{2z, 1}}
			)	
		}
	]
	-
	\Mean_{\lambda}[
		\frac{\lambda}{
			\hat{\chi}_{\Sf{2x}}
			+
			\lambda
			\hat{\chi}_{\Sf{2z}}
		}
	]
)
$
\\
$
H_{\Sf{z, k}}
=
\frac{1}{ \alpha L_{k} }
(
	\Mean_{\lambda}[
		\frac{\lambda}{
			(
				\hat{\chi}_{\Sf{2x}}
				-
				\sum_{i = 1}^{k}{}
				L_{i} \hat{H}_{\Sf{2x, i}}
			)
			+
			\lambda
			(
				\hat{\chi}_{\Sf{2z}}
				-
				\sum_{i = 1}^{k}{}
				L_{i} \hat{H}_{\Sf{2z, i}}
			)	
		}
	]
	-
	\Mean_{\lambda}[
		\frac{\lambda}{
			(
				\hat{\chi}_{\Sf{2x}}
				-
				\sum_{i = 1}^{k - 1}{}
				L_{i} \hat{H}_{\Sf{2x, i}}
			)
			+
			\lambda
			(
				\hat{\chi}_{\Sf{2z}}
				-
				\sum_{i = 1}^{k - 1}{}
				L_{i} \hat{H}_{\Sf{2z, i}}
			)	
		}
	]
)
, \,
k \in [ 2, K ]
$
\\
$
D_{\Sf{x}}
=
\int{
	\dd \mu_{\Sf{x}}
	\dd x_{0}
} \,
a_{\Sf{x}}
x_{0}
\langle
	\cdots
	\langle
		\tilde{x}
	\rangle_{\Sf{x, 0}}
	\cdots
\rangle_{\Sf{x, K}}
$
\\
$
F_{\Sf{x}}
=
\int{
	\dd \mu_{\Sf{x}}
	\dd x_{0}
} \,
a_{\Sf{x}}
\langle
	\cdots
	\langle
		\tilde{x}
	\rangle_{\Sf{x, 0}}
	\cdots
\rangle_{\Sf{x, K}}^{2}
$
\\
$
F_{\Sf{x}}
+
\frac{1}{
	\beta
}
\chi_{\Sf{x}}
+
\sum_{k = 1}^{K}{}
H_{\Sf{x, k}}
=
\int{
	\dd \mu_{\Sf{x}}
	\dd x_{0}
} \,
a_{\Sf{x}}
\langle
	\cdots
	\langle
		\tilde{x}^{2}
	\rangle_{\Sf{x, 0}}
	\cdots
\rangle_{\Sf{x, K}}
$
\\
$
F_{\Sf{x}}
+
\sum_{i = 1}^{k}{}
H_{\Sf{x, i}}
=
\int{
	\dd \mu_{\Sf{x}}
	\dd x_{0}
} \,
a_{\Sf{x}}
\langle
	\cdots
	\langle
		\cdots
		\langle
			\tilde{x}
		\rangle_{\Sf{x, 0}}
		\cdots
	\rangle_{\Sf{x, K - k}}^{2}
	\cdots
\rangle_{\Sf{x, K}}
, \,
k \in [ 1, K ]
$
\\
$
D_{\Sf{z}}
=
\int{
	\dd y
	\dd z_{0}
	\dd \mu_{\Sf{z}}
} \,
a_{\Sf{z}}
z_{0}
\langle
	\cdots
	\langle
		\tilde{z}
	\rangle_{\Sf{z, 0}}
	\cdots
\rangle_{\Sf{z, K}}
$
\\
$
F_{\Sf{z}}
=
\int{
	\dd y
	\dd z_{0}
	\dd \mu_{\Sf{z}}
} \,
a_{\Sf{z}}
\langle
	\cdots
	\langle
		\tilde{z}
	\rangle_{\Sf{z, 0}}
	\cdots
\rangle_{\Sf{z, K}}^{2}
$
\\
$
F_{\Sf{z}}
+
\frac{1}{
	\beta
}
\chi_{\Sf{z}}
+
\sum_{k = 1}^{K}{}
H_{\Sf{z, k}}
=
\int{
	\dd y
	\dd z_{0}
	\dd \mu_{\Sf{z}}
} \,
a_{\Sf{z}}
\langle
	\cdots
	\langle
		\tilde{z}^{2}
	\rangle_{\Sf{z, 0}}
	\cdots
\rangle_{\Sf{z, K}}
$
\\
$
F_{\Sf{z}}
+
\sum_{i = 1}^{k}{}
H_{\Sf{z, i}}
=
\int{
	\dd y
	\dd z_{0}
	\dd \mu_{\Sf{z}}
} \,
a_{\Sf{z}}
\langle
	\cdots
	\langle
		\cdots
		\langle
			\tilde{z}
		\rangle_{\Sf{z, 0}}
		\cdots
	\rangle_{\Sf{z, K - k}}^{2}
	\cdots
\rangle_{\Sf{z, K}}
, \,
k \in [ 1, K ]
$
\end{algorithmic}
\end{algorithm}

\section{The KVASP Algorithm}

In this section, we develop a new algorithm using message passing on a replicated factor graph.

\subsection{Definition of the Generalized Vector Survey}
Let
$
\vec{\bs{x}}
\triangleq
[
	x_{1},
	\cdots,
	x_{ \tilde{L}_{K} }
]
\in
\Real^{ 1 \times \tilde{L}_{K} }
$
be a replicated row vector of the scalar
$ x_{ 0 } $.
The generalized scalar survey for
$
\vec{\bs{x}}
$
is defined as:
\begin{align}
p( \vec{\bs{x}} )
& \triangleq
\int{
	\dd m_{\Sf{x, K - 1}}
} \,
\Normal[
	m_{\Sf{x, K - 1}} | \mu_{\Sf{x}},
	v_{\Sf{x, K}}
]
\prod_{
	i_{ K - 1 } = 1
}^{
	\frac{
		L_{K}
	}{
		L_{ K - 1 }
	}
}
\int{
	\dd m_{\Sf{x, K - 2}}
} \,
\Normal[
	m_{\Sf{x, K - 2}} | m_{\Sf{x, K - 1 }},
	v_{\Sf{x, K - 1}}
]
\times
\nonumber \\
& \quad
\prod_{
	i_{ K - 2 } = 1
}^{
	\frac{
		L_{ K - 1 }
	}{
		L_{ K - 2 }
	}
}
\int{
	\dd m_{\Sf{x, K - 3}}
} \,
\Normal[
	m_{\Sf{x, K - 3}} | m_{\Sf{x, K - 2 }},
	v_{\Sf{x, K - 2}}
]
\cdots
\prod_{
	i_{1} = 1
}^{
	\frac{
		L_{2}
	}{
		L_{1}
	}
}
\int{
	\dd m_{\Sf{x, 0}}
} \,
\Normal[
	m_{\Sf{x, 0}} | m_{\Sf{x, 1}},
	v_{\Sf{x, 1}}
]
\times
\nonumber \\
& \quad
\prod_{
	i_{0} = 1
}^{
	\tilde{L}_{1}
}
\Normal[
	x_{
		\sum_{k = 1}^{K - 1}{}
		\tilde{L}_{k}
		(
			i_{k} - 1
		)
		+
		i_{0}
	}
	| m_{\Sf{x, 0}},
	\frac{1}{\beta}
	v_{\Sf{x, 0}}
]
\label{eq:Gaussian_smoothing1} \\
& =
\Normal[
	\vec{\bs{x}}
	| \mu_{\Sf{x}}
	\vec{\bs{1}}_{ \tilde{L}_{K} },
	\frac{1}{\beta}
	v_{\Sf{x, 0}}
	\Eye_{ \tilde{L}_{K} }
	+
	v_{\Sf{x, K}}
	\mathbb{1}_{ \tilde{L}_{K} }
	+
	\sum_{k = 1}^{K - 1}{}
	v_{\Sf{x, k}}
	\Eye_{
		\frac{
			L_{K}
		}{
			L_{k}
		}
	}
	\otimes
	\mathbb{1}_{
		\tilde{L}_{k}
	}
]
, \label{Eq:Def_VS:Gau_Form}
\end{align}
where
$ \tilde{L}_{K} $
is the number of copies.
In the context of the survey propagation algorithm, a survey refers to a probabilistic measure that captures the distribution of beliefs or messages across a vast array of possible solutions to an inference problem.
In scenarios where there are multiple, often exponentially many solutions, a survey reflects the probability distribution of potential messages exchanged between nodes (or variables) in a factor graph.

Here, we extend the concept of a generalized survey for a scalar variable by defining a replicated matrix
$ \bs{X} $
for the vector
$ \bs{x}_{0} $
as follows:
\begin{align*}
\bs{X}
& \triangleq
[
	\vec{\bs{x}}_{1}^{\Ts},
	\cdots,
	\vec{\bs{x}}_{N}^{\Ts}
]^{\Ts}
\in
\Real^{ N \times \tilde{L}_{K} }
.
\end{align*}
The corresponding generalized vector survey is then defined:
\begin{align*}
p(\bs{X})
& \triangleq
\prod_{i = 1}^{N}
p( \vec{\bs{x}}_{i} )
=
\Normal[
	\tilde{\bs{x}}
	|
	\bs{1}_{ \tilde{L}_{K} }
	\otimes
	\bs{\mu}_{\Sf{x}},
	\Delta(
		\frac{1}{ \beta }
		\bs{v}_{\Sf{0}},
		\{
			\bs{v}_{\Sf{k}}
		\}_{k = 1}^{K}
	)
]
,
\end{align*}
where
$
\tilde{\bs{x}}
\triangleq
\text{vec}(\bs{X})
$
and
$
\Delta(
	\frac{1}{ \beta }
	\bs{v}_{\Sf{0}},
	\{
		\bs{v}_{\Sf{k}}
	\}_{k = 1}^{K}
)
\triangleq
\Eye_{ \tilde{L}_{K} }
\otimes
\Diag(
	\frac{1}{\beta}
	\bs{v}_{\Sf{x, 0}}
)
+
\mathbb{1}_{ \tilde{L}_{K} }
\otimes
\Diag(
	\bs{v}_{\Sf{x, K}}
)
+
\sum_{k = 1}^{K - 1}{}
\Eye_{
	\frac{
		L_{K}
	}{
		L_{k}
	}
}
\otimes
\mathbb{1}_{
	\tilde{L}_{k}
}
\otimes
\Diag(
	\bs{v}_{\Sf{x, k}}
)
$
with
$ \otimes $
denoting the Hadamard product.
Similarly, we define the generalized vector survey for
$
\vec{\bs{z}}_{m}
\triangleq
\vec{\bs{h}}_{m} \bs{X}
\in
\Real^{ 1 \times \tilde{L}_{K} }
$
with
$
\vec{\bs{h}}_{m}
$
given as
$ m $-th row vector of
$ \bs{H} $,
$
\bs{Z}
\triangleq
[
	\vec{\bs{z}}_{1}^{\Ts},
	\cdots,
	\vec{\bs{z}}_{M}^{\Ts}
]^{\Ts}
\in
\Real^{ M \times \tilde{L}_{K} }
$,
and vectorize
$\bs{Z}$
into to a column vector:
$
\tilde{\bs{z}}
\triangleq
\text{vec}(\bs{Z})
=
\tilde{\bs{H}}
\tilde{\bs{x}}
$,
where
$
\tilde{\bs{H}}
\triangleq
\Eye_{ \tilde{L}_{K} }
\otimes
\bs{H}
$.

\subsection{Propagation of the Generalized Vector Survey}

\begin{figure}[!t]
\centering
\includegraphics[width=0.85\textwidth]{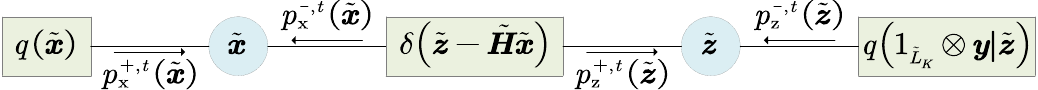}
\caption{Replicated factor graph for the inference problem}
\label{Fig:VASP}
\end{figure}

With the introduction of generalized vector surveys, we now present a replicated factor graph and propagate the surveys across it.
Fig. \ref{Fig:VASP} illustrates the replicated factor graph, where messages sent from left to right are indicated with a superscript ``$+$'', while those sent in the opposite direction are marked with a superscript ``$-$''.

Next, we iteratively update these surveys in a manner similar to VAMP and its variants \cite{zou2018concise,pandit2020inference,chen2023vector,rangan2019vector,he2017generalized}:
\begin{itemize}

\item
Backward propagation:
\begin{align}
p_{\Sf{z}}^{-, t}( \tilde{\bs{z}} )
& \propto
\frac{
	\Proj[
		q^{ \beta }(
			\bs{1}_{ \tilde{L}_{K} }
			\otimes
			\bs{y} | \tilde{\bs{z}}
		)
		p_{\Sf{z}}^{+, t}( \tilde{\bs{z}} )
	]
}{
	p_{\Sf{z}}^{+, t}( \tilde{\bs{z}} )
}
, \label{Eq:VSP:P_sz} \\
f_{\Sf{x}}^{-, t}( \tilde{\bs{x}} )
& \triangleq
\int{ \dd \tilde{\bs{z}} } \,
\delta(
	\tilde{\bs{z}}
	-
	\tilde{\bs{H}} \tilde{\bs{x}}
)
p_{\Sf{z}}^{-, t}( \tilde{\bs{z}} )
, \label{Eq:VSP:f_sx} \\
p_{\Sf{x}}^{-, t}( \tilde{\bs{x}} )
& \propto
\frac{
	\Proj[
		f_{\Sf{x}}^{-, t}( \tilde{\bs{x}} )
		p_{\Sf{x}}^{+, t}( \tilde{\bs{x}} )
	]
}{
	p_{\Sf{x}}^{+, t}( \tilde{\bs{x}} )
}
; \label{Eq:VSP:P_sx}
\end{align}

\item
Forward propagation:
\begin{align}
p_{\Sf{x}}^{+, t + 1}( \tilde{\bs{x}} )
& \propto
\frac{
	\Proj[
		q^{ \beta }( \tilde{\bs{x}} )
		p_{\Sf{x}}^{-, t}( \tilde{\bs{x}} )
	]
}{
	p_{\Sf{x}}^{-, t}( \tilde{\bs{x}} )
}
, \label{Eq:VSP:P_px} \\
f_{\Sf{z}}^{+, t + 1}( \tilde{\bs{z}} )
& \triangleq
\int{ \dd \tilde{\bs{x}} } \,
\delta(
	\tilde{\bs{z}}
	-
	\tilde{\bs{H}} \tilde{\bs{x}}
)
p_{\Sf{x}}^{+, t + 1}( \tilde{\bs{x}} )
, \label{Eq:VSP:f_pz} \\
p_{\Sf{z}}^{+, t + 1}( \tilde{\bs{z}} )
& \propto
\frac{
	\Proj[
		f_{\Sf{z}}^{+, t + 1}(
			\tilde{\bs{z}}
		)
		p_{\Sf{z}}^{-, t}( \tilde{\bs{z}} )
	]
}{
	p_{\Sf{z}}^{-, t}( \tilde{\bs{z}} )
}
; \label{Eq:VSP:P_pz}
\end{align}

\item
Iteration:
Then, back to the backward propagation and continue iterating until a stopping criterion is met.
\end{itemize}

The projection
$
\Proj[
	g( \tilde{\bs{x}} )
]
$
is a key component of the propagation process.
Its goal is to identify the multivariate Gaussian distribution that is closest to any positive function
$
g( \tilde{\bs{x}} )
$
by minimizing the Kullback-Leibler divergence (KLD).
Formally, it is defined as:
$
\Proj[
	g( \tilde{\bs{x}} )
]
\triangleq
\ArgMin\limits_{ h \in \Normal }
\mathrm{KL}[ g \| h ]	
$,
where the KLD is given by:
$
\mathrm{KL}[ g \| h ]
\triangleq
\int{ \dd \tilde{\bs{x}} } \,
g( \tilde{\bs{x}} )
\log
\frac{
	g( \tilde{\bs{x}} )
}{
	h( \tilde{\bs{x}} )
}
$.
The projection is accomplished by matching moments \cite{bishop2006pattern,minka2001family,opper2005expectation}, leading to:
\begin{align*}
\Proj[g( \tilde{\bs{x}} )]
& =
\Normal[
	\tilde{\bs{x}} |
	\bs{1}_{ \tilde{L}_{K} }
	\otimes
	\bs{m}_{ \Sf{g} },
	\Delta(
		\frac{1}{ \beta }
		\bs{v}_{ \Sf{g, 0} },
		\{
			\bs{v}_{ \Sf{g, k} }
		\}_{k = 1}^{K}
	)
]	
, \\
\bs{1}_{ \tilde{L}_{K} }
\otimes
\bs{m}_{ \Sf{g} }
& =
\frac{
	\int{ \dd \tilde{\bs{x}} } \,
	g( \tilde{\bs{x}} )
	\tilde{\bs{x}}
}{
	\int{ \dd \tilde{\bs{x}} } \,
	g( \tilde{\bs{x}} )
}
, \\
\Delta(
	\frac{1}{ \beta }
	\bs{v}_{ \Sf{g, 0} },
	\{
		\bs{v}_{ \Sf{g, k} }
	\}_{k = 1}^{K}
)
& =
\mathbb{d}[
	\frac{
		\int{ \dd \tilde{\bs{x}} } \,
		g( \tilde{\bs{x}} )
		(
			\tilde{\bs{x}}
			-
			\bs{1}_{ \tilde{L}_{K} }
			\otimes
			\bs{m}_{ \Sf{g} }
		)
		(
			\tilde{\bs{x}}
			-
			\bs{1}_{ \tilde{L}_{K} }
			\otimes
			\bs{m}_{ \Sf{g} }
		)^{\Ts}
	}{
		\int{ \dd \tilde{\bs{x}} } \,
		g( \tilde{\bs{x}} )
	}
]
,
\end{align*}
where
$
\mathbb{d}(\bs{A})
\triangleq
[
	\diag(
		\bs{A}^{(i, j)}
	)
]
$
with
$
\bs{A}
\triangleq
[
	\bs{A}^{ ( i, j ) }
]
$,
$
\bs{A}^{ ( i, j ) }
\in
\Real^{ N \times N }
$,
$
i
\in
[
	1, L_{K}
]
$,
$
j
\in
[
	1, L_{K}
]
$.
Next, we explicitly compute \eqref{Eq:VSP:P_sz}-\eqref{Eq:VSP:P_pz}\footnote{
	For symmetric matrices
	$ \bs{A} $,
	$ \bs{B} $,
	and
	$
	\bs{Q}
	\triangleq
	\mathbb{1}_{L}
	\otimes
	\bs{B}
	+
	\Eye_{L}
	\otimes
	(
		\bs{A} - \bs{B}
	)
	$,
	it holds
	$
	\bs{Q}^{-1}
	=
	-
	\mathbb{1}_{L}
	\otimes
	[
		(
			\bs{A} - \bs{B}
		)
		\bs{B}^{-1}
		(
			\bs{A} - \bs{B}
		)
		+
		L
		(
			\bs{A} - \bs{B}
		)
	]^{-1}
	+
	\Eye_{L}
	\otimes
	(
		\bs{A} - \bs{B}
	)^{- 1}
	$.
}.
With all surveys explicitly computed, we can now iteratively schedule them to develop a new survey propagation algorithm.
We propose the KVASP algorithm, detailed in Algo. \ref{Tab:VASP}, for estimating
$ \bs{x}_{0} $.
The KVASP algorithm includes VAMP \cite{rangan2019vector} as a special case, by setting
$
\{
	v_{ \Sf{x, k} }^{-}
\}_{ k = 1}^{K}
\to 0
$.
Notably, KVASP maintains the same computational complexity as VAMP, which is
$ O( N^{3} ) $
per iteration, primarily due to matrix inversion as detailed in Lines 11 and 23 of Algo. \ref{Tab:VASP}.
As noted in VAMP \cite{rangan2019vector}, if the SVD of
$ \bs{H} $
is known, this complexity can be further reduced to
$ O(N^2) $.

\begin{algorithm}[!t]
\caption{The KVASP Algorithm}
\label{Tab:VASP}
\scriptsize
\DontPrintSemicolon
\SetAlgoLined
\TB{Input:}
$
\bs{y}
$,
$
\bs{H}
$,
$
q( \bs{x}_{0} )
$,
$
q( \bs{y} | \bs{z}_{0} )
$,
$
\{
	L_{k}
\}_{k = 1}^{K}
$
\;
\TB{Initialize:}
$
\bs{\mu}_{\Sf{z}}^{+, 1}
$,
$
\{
	\bs{c}_{\Sf{z, k}}^{+, 1}
\}_{k = 0}^{K}
$,
$
\bs{\mu}_{\Sf{x}}^{+, 1}
$,
$
\{
	\bs{c}_{\Sf{x, k}}^{+, 1}
\}_{k = 0}^{K}
$
\;
\For{
	$ t \leftarrow 1 $ \KwTo $T$
}{
	$
	\bs{v}_{\Sf{z, 0}}^{+, t}
	=
	\bs{c}_{\Sf{z, 0}}^{+, t}
	$
	\;
	$
	\bs{v}_{\Sf{z, k}}^{+, t}
	=
	\frac{1}{ L_{k} }
	(
		\bs{c}_{\Sf{z, k}}^{+, t}
		-
		\bs{c}_{\Sf{z, k - 1}}^{+, t}
	)
	, \,
	k \in [ 1, K ]
	$
	\;
	$
	(
		\hat{\bs{\mu}}_{\Sf{z}}^{-, t},
		\{
			\hat{\bs{v}}_{\Sf{z, k}}^{-, t}
		\}_{k = 0}^{K}
	)
	=
	\Mean\Var[
		\bs{z} |
		\bs{\mu}_{\Sf{z}}^{+, t},
		\{
			\bs{v}_{\Sf{z, k}}^{+, t}
		\}_{k = 0}^{K},
		\{
			L_{k}
		\}_{k = 1}^{K}
	]
	$
	\;
	$
	\hat{\bs{c}}_{\Sf{z, 0}}^{-, t}
	=
	\hat{\bs{v}}_{\Sf{z, 0}}^{-, t}
	$
	\;
	$
	\hat{\bs{c}}_{\Sf{z, k}}^{-, t}
	=
	\hat{\bs{c}}_{\Sf{z, k - 1}}^{-, t}
	+
	L_{k}
	\hat{\bs{v}}_{\Sf{z, k}}^{-, t}
	, \,
	k \in [ 1, K ]
	$
	\;
	$
	\bs{c}_{\Sf{z, k}}^{-, t}
	=
	\bs{1} \oslash
	(
		\bs{1} \oslash
		\hat{\bs{c}}_{\Sf{z, k}}^{-, t}
		-
		\bs{1} \oslash
		\bs{c}_{\Sf{z, k}}^{+, t}
	)
	, \,
	k \in [ 0, K ]
	$
	\;
	$
	\bs{\mu}_{\Sf{z}}^{-, t}
	=
	\bs{c}_{\Sf{z, K}}^{-, t}
	\odot
	(
		\hat{\bs{\mu}}_{\Sf{z}}^{-, t}
		\oslash
		\hat{\bs{c}}_{\Sf{z, K}}^{-, t}
		-
		\bs{\mu}_{\Sf{z}}^{+, t}
		\oslash
		\bs{c}_{\Sf{z, K}}^{+, t}
	)
	$
	\;
	$
	\hat{\bs{C}}_{\Sf{x, k}}^{-, t}
	=
	[
		\Diag(
			\bs{1} \oslash
			\bs{c}_{\Sf{x, k}}^{+, t}
		)
		+
		\bs{H}^{\Ts}
		\Diag(
			\bs{1} \oslash
			\bs{c}_{\Sf{z, k}}^{-, t}
		)
		\bs{H}
	]^{-1}
	, \,
	k \in [ 0, K ]
	$
	\;
	$
	\hat{\bs{\mu}}_{\Sf{x}}^{-, t}
	=
	\hat{\bs{C}}_{\Sf{x, K}}^{-, t}
	[
		\bs{\mu}_{\Sf{x}}^{+, t}
		\oslash
		\bs{c}_{\Sf{x, K}}^{+, t}
		+ \bs{H}^{\Ts}
		(
			\bs{\mu}_{\Sf{z}}^{-, t}
			\oslash
			\bs{c}_{\Sf{z, K}}^{-, t}
		)
	]
	$
	\;
	$
	\hat{\bs{c}}_{\Sf{x, k}}^{-, t}
	=
	\Sf{d}(
		\hat{\bs{C}}_{\Sf{x, k}}^{-, t}
	)
	, \,
	k \in [ 0, K ]
	$
	\;
	$
	\bs{c}_{\Sf{x, k}}^{-, t}
	=
	\bs{1} \oslash
	(
		\bs{1}
		\oslash
		\hat{\bs{c}}_{\Sf{x, k}}^{-, t}
		-
		\bs{1}
		\oslash
		\bs{c}_{\Sf{x, k}}^{+, t}
	)
	, \,
	k \in [ 0, K ]
	$
	\;
	$
	\bs{\mu}_{\Sf{x}}^{-, t}
	=
	\bs{c}_{\Sf{x, K}}^{-, t}
	\odot
	(
		\hat{\bs{\mu}}_{\Sf{x}}^{-, t}
		\oslash
		\hat{\bs{c}}_{\Sf{x, K}}^{-, t}
		-
		\bs{\mu}_{\Sf{x}}^{+, t}
		\oslash
		\bs{c}_{\Sf{x, K}}^{+, t}
	)
	$
	\;
	$
	\bs{v}_{\Sf{x, 0}}^{-, t}
	=
	\bs{c}_{\Sf{x, 0}}^{-, t}
	$
	\;
	$
	\bs{v}_{\Sf{x, k}}^{-, t}
	=
	\frac{1}{ L_{k} }
	(
		\bs{c}_{\Sf{x, k}}^{-, t}
		-
		\bs{c}_{\Sf{x, k - 1}}^{-, t}
	)
	, \,
	k \in [ 1, K ]
	$
	\;	
	$
	(
		\hat{\bs{\mu}}_{\Sf{x}}^{+, t + 1},
		\{
			\hat{\bs{v}}_{\Sf{x, k}}^{+, t + 1}
		\}_{k = 0}^{K}
	)
	=
	\Mean\Var[
		\bs{x} |
		\bs{\mu}_{\Sf{x}}^{-, t},
		\{
			\bs{v}_{\Sf{x, k}}^{-, t}
		\}_{k = 0}^{K},
		\{
			L_{k}
		\}_{k = 1}^{K}
	]
	$
	\;
	$
	\hat{\bs{c}}_{\Sf{x, 0}}^{+, t + 1}
	=
	\hat{\bs{v}}_{\Sf{x, 0}}^{+, t + 1}
	$
	\;
	$
	\hat{\bs{c}}_{\Sf{x, k}}^{+, t + 1}
	=
	\hat{\bs{c}}_{\Sf{x, k - 1}}^{+, t + 1}
	+
	L_{k}
	\hat{\bs{v}}_{\Sf{x, k}}^{+, t + 1}
	, \,
	k \in [ 1, K ]
	$
	\;
	$
	\bs{c}_{\Sf{x, k}}^{+, t + 1}
	=
	\bs{1}
	\oslash
	(
		\bs{1}
		\oslash
		\hat{\bs{c}}_{\Sf{x, k}}^{+, t + 1}
		-
		\bs{1}
		\oslash
		\bs{c}_{\Sf{x, k}}^{-, t}
	)
	, \,
	k \in [ 0, K ]
	$
	\;
	$
	\bs{\mu}_{\Sf{x}}^{+, t + 1}
	=
	\bs{c}_{\Sf{x, K}}^{+, t + 1}
	\odot
	(
		\hat{\bs{\mu}}_{\Sf{x}}^{+, t + 1} \oslash
		\hat{\bs{c}}_{\Sf{x, K}}^{+, t + 1}
		-
		\bs{\mu}_{\Sf{x}}^{-, t}
		\oslash
		\bs{c}_{\Sf{x, K}}^{-, t}
	)
	$
	\;
	$
	\hat{\bs{C}}_{\Sf{\tilde{x}, k}}^{+, t + 1}
	=
	[
		\Diag(
			\bs{1}
			\oslash
			\bs{c}_{\Sf{x, k}}^{+, t + 1}
		)
		+
		\bs{H}^{\Ts}
		\Diag(
			\bs{1}
			\oslash
			\bs{c}_{\Sf{z, k}}^{-, t}
		)
		\bs{H}
	]^{-1}
	, \,
	k \in [ 0, K ]
	$
	\;
	$
	\hat{\bs{\mu}}_{\Sf{\tilde{x}}}^{+, t + 1}
	=
	\hat{\bs{C}}_{\Sf{\tilde{x}, K}}^{+, t + 1}
	[
		\bs{\mu}_{\Sf{x}}^{+, t + 1}
		\oslash
		\bs{c}_{\Sf{x, K}}^{+, t + 1}
		+
		\bs{H}^{\Ts}
		(
			\bs{\mu}_{\Sf{z}}^{-, t}
			\oslash
			\bs{c}_{\Sf{z, K}}^{-, t}
		)
	]
	$
	\;
	$
	\hat{\bs{C}}_{\Sf{z, k}}^{+, t + 1}
	=
	\bs{H}
	\hat{\bs{C}}_{\Sf{\tilde{x}, k}}^{+, t + 1}
	\bs{H}^{\Ts}
	, \,
	k \in [ 0, K ]
	$
	\;
	$
	\hat{\bs{\mu}}_{\Sf{z}}^{+, t + 1}
	=
	\bs{H}
	\hat{\bs{\mu}}_{\Sf{\tilde{x}}}^{+, t + 1}
	$
	\;
	$
	\hat{\bs{c}}_{\Sf{z, k}}^{+, t + 1}
	=
	\Sf{d}(
		\hat{\bs{C}}_{\Sf{z, k}}^{+, t + 1}
	)
	, \,
	k \in [ 0, K ]
	$
	\;
	$
	\bs{c}_{\Sf{z, k}}^{+, t + 1}
	=
	\bs{1} \oslash
	(
		\bs{1}
		\oslash
		\hat{\bs{c}}_{\Sf{z, k}}^{+, t + 1}
		-
		\bs{1}
		\oslash
		\bs{c}_{\Sf{z, k}}^{-, t}
	)
	, \,
	k \in [ 0, K ]
	$
	\;
	$
	\bs{\mu}_{\Sf{z}}^{+, t + 1}
	=
	\bs{c}_{\Sf{z, K}}^{+, t + 1}
	\odot
	(
		\hat{\bs{\mu}}_{\Sf{z}}^{+, t + 1} \oslash
		\hat{\bs{c}}_{\Sf{z, K}}^{+, t + 1}
		-
		\bs{\mu}_{\Sf{z}}^{-, t}
		\oslash
		\bs{c}_{\Sf{z, K}}^{-, t}
	)
	$
	\;
}
\TB{Output:}
$
\hat{\bs{\mu}}_{\Sf{x}}^{+, T + 1}
$,
$
\{
	\hat{\bs{v}}_{\Sf{x, k}}^{+, T + 1}
\}_{k = 0}^{K}
$
\end{algorithm}

\section{State Evolution of the KVASP}

\subsection{SE and Its Fixed Point Equations}

In this subsection, we introduce a set of SE equations to capture the dynamics of KVASP.
The SE is based on an assumption derived from two prior studies on mismatched problems \cite{gerbelot2022asymptotic,takahashi2022macroscopic}.
This assumption posits that the parameters of the input and output to the endpoint factor nodes in Fig. \ref{Fig:VASP}--specifically,
$
\bs{\mu}_{\Sf{x}}^{-, t}
$,
$
\bs{\mu}_{\Sf{z}}^{+, t}
$,
$
\bs{\mu}_{\Sf{x}}^{+, t}
$,
$
\bs{\mu}_{\Sf{z}}^{-, t}
$--all converge to Gaussian random variables centered around
$ \bs{x}_{0} $
and
$ \bs{z}_{0} $,
subject to some rotations.

\begin{assumption}[Empirical Convergence \cite{gerbelot2022asymptotic,takahashi2022macroscopic}]\label{assump:EmpiricalConverge}

At each iteration, the intermediate vectors
$
\bs{\mu}_{\Sf{x}}^{+, t}
$,
$
\bs{\mu}_{\Sf{x}}^{-, t}
$,
$
\bs{\mu}_{\Sf{z}}^{+, t}
$,
$
\bs{\mu}_{\Sf{z}}^{-, t}
$
empirically converge with second-order moment (PL2) towards Gaussian variables as described in LSL
\begin{align}
\lim\limits_{
	M, N \rightarrow \infty
}
\frac{1}{
	\mathtt{c}_{\Sf{x, K}}^{-, t}
}
\bs{\mu}_{\Sf{x}}^{-, t}
-
\hat{D}_{\Sf{1x}}^{t}
\bs{x}_{0}
& \overset{ \text{PL2} }{=}
\sqrt{
	\hat{F}_{\Sf{1x}}^{t}
}
\xi_{\Sf{1x}}^{t}
, \tag{a1} \\
\lim\limits_{
	M, N \rightarrow \infty
}
\frac{1}{
	\mathtt{c}_{\Sf{x, K}}^{+, t}
}
\bs{V}^{\Ts}
\bs{\mu}_{\Sf{x}}^{+, t}
-
\hat{D}_{\Sf{2x}}^{t}
\bs{V}^{\Ts} \bs{x}_{0}
& \overset{ \text{PL2} }{=}
\sqrt{
	\hat{F}_{\Sf{2x}}^{t}
}
\xi_{\Sf{2x}}^{t}
, \tag{a2} \\
\lim\limits_{
	M, N \rightarrow \infty
}
\frac{1}{
	\mathtt{c}_{\Sf{z, K}}^{-, t}
}
\bs{\mu}_{\Sf{z}}^{-, t}
-
\hat{D}_{\Sf{2z}}^{t}
\bs{z}_{0}
& \overset{ \text{PL2} }{=}
\sqrt{
	\hat{F}_{\Sf{2z}}^{t}
}
\xi_{\Sf{2z}}^{t}
, \tag{a3} \\
\lim\limits_{
	M, N \rightarrow \infty
}
\frac{1}{
	\mathtt{c}_{\Sf{z, K}}^{+, t}
}
\bs{U}^{\Ts}
\bs{\mu}_{\Sf{z}}^{+, t}
-
\hat{D}_{\Sf{1z}}^{t}
\bs{U}^{\Ts} \bs{z}_{0}
& \overset{ \text{PL2} }{=}
\sqrt{
	\hat{F}_{\Sf{1z}}^{t}
}
\xi_{\Sf{1z}}^{t}
, \tag{a4}
\end{align}
where
$ \bs{U} $
and
$ \bs{V} $
are obtained from the SVD of
$
\bs{H}
=
\bs{U} \bs{S} \bs{V}^{\Ts}
$,
and the random variables
$ \{\xi\} $
are i.i.d. standard Gaussian, the coefficients are defined as follows:
$
\hat{D}_{\Sf{1x}}^{t}
=
\Mean[
	\frac{1}{
		N C_{\Sf{x}}
		\mathtt{c}_{\Sf{x, K}}^{-, t}
	}
	\bs{x}_{0}^{\Ts}
	\bs{\mu}_{\Sf{x}}^{-, t}
]
$,
$
\hat{F}_{\Sf{1x}}^{t}
=
\Mean[
	\frac{1}{
		N
		(
			\mathtt{c}_{\Sf{x, K}}^{-, t}
		)^{2}
	}
	\|
		\bs{\mu}_{\Sf{x}}^{-, t}
		-
		\mathtt{c}_{\Sf{x, K}}^{-, t}
		\hat{D}_{\Sf{1x}}^{t}
		\bs{x}_{0}
	\|_{2}^{2}
]
$,
$
\hat{D}_{\Sf{2x}}^{t}
=
\Mean[
	\frac{1}{
		N C_{\Sf{x}}
		\mathtt{c}_{\Sf{x, K}}^{+, t}
	}
	\bs{x}_{0}^{\Ts}
	\bs{\mu}_{\Sf{x}}^{+, t}
]
$,
$
\hat{F}_{\Sf{2x}}^{t}
=
\Mean[
	\frac{1}{
		N
		(
			\mathtt{c}_{\Sf{x, K}}^{+, t}
		)^{2}
	}
	\|
		\bs{\mu}_{\Sf{x}}^{+, t}
		-
		\mathtt{c}_{\Sf{x, K}}^{+, t}
		\hat{D}_{\Sf{2x}}^{t}
		\bs{x}_{0}
	\|_{2}^{2}
]
$,
$
\hat{D}_{\Sf{2z}}^{t}
=
\Mean[
	\frac{1}{
		M C_{\Sf{z}}
		\mathtt{c}_{\Sf{z, K}}^{-, t}
	}
	\bs{z}_{0}^{\Ts}
	\bs{\mu}_{\Sf{z}}^{-, t}
]
$,
$
\hat{F}_{\Sf{2z}}^{t}
=
\Mean[
	\frac{1}{
		M
		(
			\mathtt{c}_{\Sf{z, K}}^{-, t}
		)^{2}
	}
	\|
		\bs{\mu}_{\Sf{z}}^{-, t}
		-
		\mathtt{c}_{\Sf{z, K}}^{-, t}
		\hat{D}_{\Sf{2z}}^{t}
		\bs{z}_{0}
	\|_{2}^{2}
]
$,
$
\hat{D}_{\Sf{1z}}^{t}
=
\Mean[
	\frac{1}{
		M C_{\Sf{z}}
		\mathtt{c}_{\Sf{z, K}}^{+, t}
	}
	\bs{z}_{0}^{\Ts}
	\bs{\mu}_{\Sf{z}}^{+, t}
]
$,
$
\hat{F}_{\Sf{1z}}^{t}
=
\Mean[
	\frac{1}{
		M
		(
			\mathtt{c}_{\Sf{z, K}}^{+, t}
		)^{2}
	}
	\|
		\bs{\mu}_{\Sf{z}}^{+, t}
		-
		\mathtt{c}_{\Sf{z}}^{+, t}
		\hat{D}_{\Sf{1z}}^{t}
		\bs{z}_{0}
	\|_{2}^{2}
]
$,
where
$
C_{\Sf{x}}
=
\Mean[ x_{0}^{2} ]
$,
$
C_{\Sf{z}}
=
\frac{1}{\alpha}
\Mean_{\lambda}[\lambda]
C_{\Sf{x}}
$,
and
$ \lambda $
follows the limiting eigenvalue distribution of
$ \bs{H}^{\Ts} \bs{H} $.
\end{assumption}

We propose the SE equations in Algo. \ref{Tab:SE} to capture the dynamics of KVASP.
Specifically, the per-iteration MSE, defined as
$
\Sf{MSE}^{t}
\triangleq
\frac{1}{C_{\Sf{x}}}
\Mean[
	\frac{1}{N}
	\|
		\hat{\bs{\mu}}_{\Sf{x}}^{+, t}
		-
		\bs{x}_{0}
	\|^{2}
]
$,
is characterized by
$
\Sf{MSE}^{t}
=
\frac{1}{C_{\Sf{x}}}
(
	C_{\Sf{x}}
	+
	F_{\Sf{x}}^{+, t}
	-
	2 D_{\Sf{x}}^{+, t}
)
$,
where the macroscopic parameters are defined as follows:
$
C_{\Sf{x}}
\triangleq
\Mean[ x_{0}^{2} ]
$,
$
D_{\Sf{x}}^{+, t}
\triangleq
\Mean[
	\frac{1}{N}
	\bs{x}_{0}^{\Ts}
	\hat{\bs{\mu}}_{\Sf{x}}^{+, t}
]
$,
$
F_{\Sf{x}}^{+, t}
\triangleq
\Mean[
	\frac{1}{N}
	\|
		\hat{\bs{\mu}}_{\Sf{x}}^{+, t}
	\|_{2}^{2}
]
$.
In the following section, we will connect the SE to the KRSB saddle point equations outlined in Algo. \ref{Tab:Saddle_Point}.

\begin{proposition} \label{Proposition:FixedPoint}
The fixed point equations of the KVASP's SE align with Algo. \ref{Tab:Saddle_Point}.
\end{proposition}

\begin{IEEEproof}
Refer to Appendix \ref{Appendix:Fixed_Point} for a sketch of the proof.
\end{IEEEproof}

This type of correspondence is frequently observed in the literature on AMP and GASP \cite{takahashi2022macroscopic, gerbelot2022asymptotic,barbier2023compressed,he2017generalized,lucibello2019generalized,antenucci2019approximate}.
As discussed in \cite{takahashi2022macroscopic}, the SE equations represent iterations of the variational conditions for free energy under the RS/RSB ansatz, with their fixed points corresponding to the saddle points of the free energy.
It is important to note that the precise number of SE fixed points in a generic AMP or ASP algorithm remains uncertain within the community.
Multiple solutions may exist for the fixed point equations, and those solutions that globally minimize the free energy are particularly significant, referred to as (globally) stable solutions \cite{guo2009generic}.
Conversely, solutions that locally minimize the free energy are termed metastable, while those that locally maximize it are labeled unstable.

The SE-to-replica correspondence implies that the proposed KVASP algorithm asymptotically achieves the ideal estimator \cite{rangan2012asymptotic} with cubic complexity, provided that the algorithm converges to an SE fixed point that is a globally stable solution of the free energy's KRSB saddle point equations.
Recall that VAMP serves as the estimator under the RS ansatz \cite{takahashi2022macroscopic}.
Our work extends VAMP to the KRSB case.

\begin{algorithm}[!t]
\caption{State Evolution of KVASP}
\label{Tab:SE}
\scriptsize
\DontPrintSemicolon
\SetAlgoLined
\TB{Input:}
$
p( \bs{x}_{0} )
$,
$
p( \bs{y} | \bs{z}_{0} )
$,
$
q( \bs{x}_{0} )
$,
$
q( \bs{y} | \bs{z}_{0} )
$,
$
\{
	L_{k}
\}_{k = 1}^{K}
$
\;
\TB{Initialize:}
$
\hat{D}_{\Sf{1z}}^{1}
$,
$
\hat{F}_{\Sf{1z}}^{1}
$,
$
\{
	\mathtt{c}_{\Sf{z, k}}^{+, 1}
\}_{k = 0}^{K}
$,
$
\hat{D}_{\Sf{2x}}^{1}
$,
$
\hat{F}_{\Sf{2x}}^{1}
$,
$
\{
	\mathtt{c}_{\Sf{x, k}}^{+, 1}
\}_{k = 0}^{K}
$
\;
\For{
	$ t \leftarrow 1 $ \KwTo $T$
}{
	$
	\mathtt{v}_{\Sf{z, 0}}^{+, t}
	=
	\mathtt{c}_{\Sf{z, 0}}^{+, t}
	$
	\;
	$
	\mathtt{v}_{\Sf{z, k}}^{+, t}
	=
	\frac{1}{ L_{k} }
	(
		\mathtt{c}_{\Sf{z, k}}^{+, t}
		-
		\mathtt{c}_{\Sf{z, k - 1}}^{+, t}
	)
	, \,
	k \in [ 1, K ]
	$
	\;
	$
	D_{\Sf{z}}^{-, t}
	=
	\int{
		\dd y \dd z_{0} \dd \mu_{\Sf{z}}
	} \,
	a^{t}_{\Sf{z}} z_{0}
	\langle
		\cdots
		\langle
			\tilde{z}
		\rangle_{\Sf{z, 0}}
		\cdots
	\rangle_{\Sf{z, K}}
	$
	\;
	$
	F_{\Sf{z}}^{-, t}
	=
	\int{
		\dd y \dd z_{0} \dd \mu_{\Sf{z}}
	} \,
	a^{t}_{\Sf{z}}
	\langle
		\cdots
		\langle
			\tilde{z}
		\rangle_{\Sf{z, 0}}
		\cdots
	\rangle_{\Sf{z, K}}^{2}
	$
	\;
	$
	F_{\Sf{z}}^{-, t}
	+
	\frac{1}{ \beta }
	\hat{\mathtt{v}}_{\Sf{z, 0}}^{-, t}
	+
	\sum_{k = 1}^{K}{}
	\hat{\mathtt{v}}_{\Sf{z, k}}^{-, t}
	=
	\int{
		\dd y \dd z_{0} \dd \mu_{\Sf{z}}
	} \,
	a^{t}_{\Sf{z}}
	\langle
		\cdots
		\langle
			\tilde{z}^{2}
		\rangle_{\Sf{z, 0}}
		\cdots
	\rangle_{\Sf{z, K}}
	$
	\;
	$
	F_{\Sf{z}}^{-, t}
	+
	\sum_{i = 1}^{k}{}
	\hat{\mathtt{v}}_{\Sf{z, i}}^{-, t}
	=
	\int{
		\dd y \dd z_{0} \dd \mu_{\Sf{z}}
	} \,
	a^{t}_{\Sf{z}}
	\langle
		\cdots
		\langle
			\cdots
			\langle
				\tilde{z}
			\rangle_{\Sf{z, 0}}
			\cdots
		\rangle_{\Sf{z, K - k}}^{2}
		\cdots
	\rangle_{\Sf{z, K}}
	, \,
	k \in [ 1, K ]
	$
	\;
	$
	\hat{\mathtt{c}}_{\Sf{z, 0}}^{-, t}
	=
	\hat{\mathtt{v}}_{\Sf{z, 0}}^{-, t}
	$
	\;
	$
	\hat{\mathtt{c}}_{\Sf{z, k}}^{-, t}
	=
	\hat{\mathtt{c}}_{\Sf{z, k - 1}}^{-, t}
	+
	L_{k} \hat{\mathtt{v}}_{\Sf{z, k}}^{-, t}
	, \,
	k \in [ 1, K ]
	$
	\;
	$
	\mathtt{c}_{\Sf{z, k}}^{-, t}
	=
	(
		\frac{1}{
			\hat{\mathtt{c}}_{\Sf{z, k}}^{-, t}
		}
		-
		\frac{1}{
			\mathtt{c}_{\Sf{z, k}}^{+, t}
		}
	)^{-1}
	, \,
	k \in [ 0, K ]
	$
	\;
	$
	\hat{D}_{\Sf{2z}}^{t}
	=
	\frac{
		D_{\Sf{z}}^{-, t}
	}{
		C_{\Sf{z}}
		\hat{\mathtt{c}}_{\Sf{z, K}}^{-, t}
	}
	-
	\hat{D}_{\Sf{1z}}^{t}
	$
	\;
	$
	\hat{F}_{\Sf{2z}}^{t}
	=
	\frac{
		F_{\Sf{z}}^{-, t}
	}{
		(
			\hat{\mathtt{c}}_{\Sf{z, K}}^{-, t}
		)^{2}
	}
	-
	\frac{
		(
			D_{\Sf{z}}^{-, t}
		)^{2}
	}{
		C_{\Sf{z}}
		(
			\hat{\mathtt{c}}_{\Sf{z, K}}^{-, t}
		)^{2}
	}
	-
	\hat{F}_{\Sf{1z}}^{t}
	$
	\;
	$
	\hat{\mathtt{c}}_{\Sf{x, k}}^{-, t}
	=
	\Mean_{\lambda}[
		(
			\frac{1}{
				\mathtt{c}_{\Sf{x, k}}^{+, t}
			}
			+
			\frac{\lambda}{
				\mathtt{c}_{\Sf{z, k}}^{-, t}
			}
		)^{-1}
	]
	, \,
	k \in [ 0, K ]
	$
	\;
	$
	D_{\Sf{x}}^{-, t}
	=
	C_{\Sf{x}}
	\Mean_{\lambda}[
		\frac{
			\hat{D}_{\Sf{2x}}^{t}
			+
			\hat{D}_{\Sf{2z}}^{t} \lambda
		}{
			\frac{1}{
				\mathtt{c}_{\Sf{x, K}}^{+, t}
			}
			+
			\frac{\lambda}{
				\mathtt{c}_{\Sf{z, K}}^{-, t}
			}
		}
	]
	$
	\;
	$
	F_{\Sf{x}}^{-, t}
	=
	C_{\Sf{x}}
	\Mean_{\lambda}[
		\frac{
			(
				\hat{D}_{\Sf{2x}}^{t}
				+
				\hat{D}_{\Sf{2z}}^{t} \lambda
			)^{2}
		}{
			(
				\frac{1}{
					\mathtt{c}_{\Sf{x, K}}^{+, t}
				}
				+
				\frac{\lambda}{
					\mathtt{c}_{\Sf{z, K}}^{-, t}
				}
			)^{2}
		}
	]
	+
	\Mean_{\lambda}[
		\frac{
			\hat{F}_{\Sf{2x}}^{t}
			+
			\hat{F}_{\Sf{2z}}^{t} \lambda
		}{
			(
				\frac{1}{
					\mathtt{c}_{\Sf{x, K}}^{+, t}
				}
				+
				\frac{\lambda}{
					\mathtt{c}_{\Sf{z, K}}^{-, t}
				}
			)^{2}
		}
	]
	$
	\;
	$
	\mathtt{c}_{\Sf{x, k}}^{-, t}
	=
	(
		\frac{1}{
			\hat{\mathtt{c}}_{\Sf{x, k}}^{-, t}
		}
		-
		\frac{1}{
			\mathtt{c}_{\Sf{x, k}}^{+, t}
		}
	)^{-1}
	, \,
	k \in [ 0, K ]
	$
	\;
	$
	\hat{D}_{\Sf{1x}}^{t}
	=
	\frac{
		D_{\Sf{x}}^{-, t}
	}{
		C_{\Sf{x}}
		\hat{\mathtt{c}}_{\Sf{x, K}}^{-, t}
	}
	-
	\hat{D}_{\Sf{2x}}^{t}
	$
	\;
	$
	\hat{F}_{\Sf{1x}}^{t}
	=
	\frac{
		F_{\Sf{x}}^{-, t}
	}{
		(
			\hat{\mathtt{c}}_{\Sf{x, K}}^{-, t}
		)^{2}
	}
	-
	\frac{
		(
			D_{\Sf{x}}^{-, t}
		)^{2}
	}{
		C_{\Sf{x}}
		(
			\hat{\mathtt{c}}_{\Sf{x, K}}^{-, t}
		)^{2}
	}
	-
	\hat{F}_{\Sf{2x}}^{t}
	$
	\;
	$
	\mathtt{v}_{\Sf{x, 0}}^{-, t}
	=
	\mathtt{c}_{\Sf{x, 0}}^{-, t}
	$
	\;
	$
	\mathtt{v}_{\Sf{x, k}}^{-, t}
	=
	\frac{1}{ L_{k} }
	(
		\mathtt{c}_{\Sf{x, k}}^{-, t}
		-
		\mathtt{c}_{\Sf{x, k - 1}}^{-, t}
	)
	, \,
	k \in [ 1, K ]
	$
	\;
	$
	D_{\Sf{x}}^{+, t + 1}
	=
	\int{
		\dd \mu_{\Sf{x}} \dd x_{0}
	} \,
	a^{t}_{\Sf{x}} x_{0}
	\langle
		\cdots
		\langle
			\tilde{x}
		\rangle_{\Sf{x, 0}}
		\cdots
	\rangle_{\Sf{x, K}}
	$
	\;
	$
	F_{\Sf{x}}^{+, t + 1}
	=
	\int{
		\dd \mu_{\Sf{x}} \dd x_{0}
	} \,
	a^{t}_{\Sf{x}}
	\langle
		\cdots
		\langle
			\tilde{x}
		\rangle_{\Sf{x, 0}}
		\cdots
	\rangle_{\Sf{x, K}}^{2}
	$
	\;
	$
	F_{\Sf{x}}^{+, t + 1}
	+
	\frac{1}{ \beta }
	\hat{\mathtt{v}}_{\Sf{x, 0}}^{+, t + 1}
	+
	\sum_{k = 1}^{K}{}
	\hat{\mathtt{v}}_{\Sf{x, k}}^{+, t + 1}
	=
	\int{
		\dd \mu_{\Sf{x}} \dd x_{0}
	} \,
	a^{t}_{\Sf{x}}
	\langle
		\cdots
		\langle
			\tilde{x}^{2}
		\rangle_{\Sf{x, 0}}
		\cdots
	\rangle_{\Sf{x, K}}
	$
	\;
	$
	F_{\Sf{x}}^{+, t + 1}
	+
	\sum_{i = 1}^{k}{}
	\hat{\mathtt{v}}_{\Sf{x, i}}^{+, t + 1}
	=
	\int{
		\dd \mu_{\Sf{x}} \dd x_{0}
	} \,
	a^{t}_{\Sf{x}}
	\langle
		\cdots
		\langle
			\cdots
			\langle
				\tilde{x}
			\rangle_{\Sf{x, 0}}
			\cdots
		\rangle_{\Sf{x, K - k}}^{2}
		\cdots
	\rangle_{\Sf{x, K}}
	, \,
	k \in [ 1, K ]
	$
	\;
	$
	\hat{\mathtt{c}}_{\Sf{x, 0}}^{+, t + 1}
	=
	\hat{\mathtt{v}}_{\Sf{x, 0}}^{+, t + 1}
	$
	\;
	$
	\hat{\mathtt{c}}_{\Sf{x, k}}^{+, t + 1}
	=
	\hat{\mathtt{c}}_{\Sf{x, k - 1}}^{+, t + 1}
	+
	L_{k}
	\hat{\mathtt{v}}_{\Sf{x, k}}^{+, t + 1}
	, \,
	k \in [ 1, K ]
	$
	\;
	$
	\mathtt{c}_{\Sf{x, k}}^{+, t + 1}
	=
	(
		\frac{1}{
			\hat{\mathtt{c}}_{\Sf{x, k}}^{+, t + 1}
		}
		-
		\frac{1}{
			\mathtt{c}_{\Sf{x, k}}^{-, t}
		}
	)^{-1}
	, \,
	k \in [ 0, K ]
	$
	\;
	$
	\hat{D}_{\Sf{2x}}^{t + 1}
	=
	\frac{
		D_{\Sf{x}}^{+, t + 1}
	}{
		C_{\Sf{x}}
		\hat{\mathtt{c}}_{\Sf{x, K}}^{+, t + 1}
	}
	-
	\hat{D}_{\Sf{1x}}^{t}
	$
	\;
	$
	\hat{F}_{\Sf{2x}}^{t + 1}
	=
	\frac{
		F_{\Sf{x}}^{+, t + 1}
	}{
		(
			\hat{\mathtt{c}}_{\Sf{x, K}}^{+, t + 1}
		)^{2}
	}
	-
	\frac{
		(
			D_{\Sf{x}}^{+, t + 1}
		)^{2}
	}{
		C_{\Sf{x}}
		(
			\hat{\mathtt{c}}_{\Sf{x, K}}^{+, t + 1}
		)^{2}
	}
	-
	\hat{F}_{\Sf{1x}}^{t}
	$
	\;
	$
	\hat{\mathtt{c}}_{\Sf{z, k}}^{+, t + 1}
	=
	\frac{1}{\alpha}
	\Mean_{\lambda}[
		\frac{\lambda}{
			(
				\frac{1}{
					\mathtt{c}_{\Sf{x, k}}^{+, t + 1}
				}
				+
				\frac{\lambda}{
					\mathtt{c}_{\Sf{z, k}}^{-, t}
				}
			)
		}
	]
	, \,
	k \in [ 0, K ]
	$
	\;
	$
	D_{\Sf{z}}^{+, t + 1}
	=
	\frac{
		C_{\Sf{x}}
	}{\alpha}
	\Mean_{\lambda}[
		\frac{
			\lambda
			(
				\hat{D}_{\Sf{2x}}^{t + 1}
				+
				\hat{D}_{\Sf{2z}}^{t} \lambda
			)
		}{
			\frac{1}{
				\mathtt{c}_{\Sf{x, K}}^{+, t + 1}
			}
			+
			\frac{\lambda}{
				\mathtt{c}_{\Sf{z, K}}^{-, t}
			}
		}
	]
	$
	\;
	$
	F_{\Sf{z}}^{+, t + 1}
	=
	\frac{
		C_{\Sf{x}}
	}{\alpha}
	\Mean_{\lambda}[
		\frac{
			\lambda (
				\hat{D}_{\Sf{2x}}^{t + 1}
				+
				\hat{D}_{\Sf{2z}}^{t} \lambda
			)^{2}
		}{
			(
				\frac{1}{
					\mathtt{c}_{\Sf{x, K}}^{+, t + 1}
				}
				+
				\frac{\lambda}{
					\mathtt{c}_{\Sf{z, K}}^{-, t}
				}
			)^{2}
		}
	]
	+
	\frac{1}{\alpha}
	\Mean_{\lambda}[
		\frac{
			\lambda
			(
				\hat{F}_{\Sf{2x}}^{t + 1}
				+
				\hat{F}_{\Sf{2z}}^{t} \lambda
			)
		}{
			(
				\frac{1}{
					\mathtt{c}_{\Sf{x, K}}^{+, t + 1}
				}
				+ \frac{\lambda}{
					\mathtt{c}_{\Sf{z, K}}^{-, t}
				}
			)^{2}
		}
	]
	$
	\;
	$
	\mathtt{c}_{\Sf{z, k}}^{+, t + 1}
	=
	(
		\frac{1}{
			\hat{\mathtt{c}}_{\Sf{z, k}}^{+, t + 1}
		}
		-
		\frac{1}{
			\mathtt{c}_{\Sf{z, k}}^{-, t}
		}
	)^{-1}
	, \,
	k \in [ 0, K ]
	$
	\;
	$
	\hat{D}_{\Sf{1z}}^{t + 1}
	=
	\frac{
		D_{\Sf{z}}^{+, t + 1}
	}{
		C_{\Sf{z}}
		\hat{\mathtt{c}}_{\Sf{z, K}}^{+, t + 1}
	}
	-
	\hat{D}_{\Sf{2z}}^{t}
	$
	\;
	$
	\hat{F}_{\Sf{1z}}^{t + 1}
	=
	\frac{
		F_{\Sf{z}}^{+, t + 1}
	}{
		(
			\hat{\mathtt{c}}_{\Sf{z, K}}^{+, t + 1}
		)^{2}
	}
	-
	\frac{
		(
			D_{\Sf{z}}^{+, t + 1}
		)^{2}
	}{
		C_{\Sf{z}}
		(
			\hat{\mathtt{c}}_{\Sf{z, K}}^{+, t + 1}
		)^{2}
	}
	-
	\hat{F}_{\Sf{2z}}^{t}
	$
	\;
}
\TB{Output:}
$
D_{\Sf{x}}^{+, T + 1}
$,
$
F_{\Sf{x}}^{+, T + 1}
$,
$
\{
	\hat{\mathtt{v}}_{\Sf{x, k}}^{+, t + 1}
\}_{k = 0}^{K}
$
\end{algorithm}

\section{Simulation Results}\label{sec:simulation}

We conduct Monte Carlo simulations to evaluate the performance of KVASP and its SE.
Our focus is on symbol detection in multiple-input multiple-output (MIMO) systems
\begin{align*}
\bs{H}
& \triangleq
\bs{H}_{\Sf{w}}
\bs{R}^{\frac{1}{2}}
, \\
\bs{z}_{0}
& \triangleq
\bs{H} \bs{x}_{0}
, \\
\bs{y}
& =
\bs{z}_{0} + \bs{w}
, \\
p( x_{0} )
& \propto
\EXP\left[
-
\frac{
	(
		| x_{0} | - 1
	)^{2}
}{
	2 c
}
\right]
, \\
p( y | z_{0} )
& =
\Normal[ y | z_{0}, v_{\Sf{T}} ]
, \\
q(x_{0})
& =
\frac{
	\delta( x_{0} + 1 )
	+
	\delta( x_{0} - 1 )
}{2}
, \\
q( y | z_{0} )
& =
\Normal[ y | z_{0}, v_{\Sf{F}} ]
,
\end{align*}
where the measurement matrix
$ \bs{H}_{\Sf{w}} $
is randomly generated from a white Gaussian ensemble, characterized by i.i.d. elements that have a zero mean and a variance of
$\frac{1}{N}$.
The deterministic correlation matrix
$ \bs{R} $
is constructed using the Kronecker correlation model \cite{paulraj2003introduction}, where the
$ ( i, j ) $-th element given by
$
R_{ i, j }
=
\rho^{ |i - j| }
$
with
$
\rho \in [ 0, 1 )
$
representing the correlation factor.
When
$ \rho = 0 $,
the correlated model simplifies to the standard i.i.d. model considered in \cite{lucibello2019generalized}.
For the postulated prior
$ q( x_{0} ) $,
we employ standard BPSK modulation.
To introduce model mismatch, the ground-truth prior
$ p( x_{0} ) $
is allowed to differ slightly;
however, as
$ c \to 0 $,
the model mismatch vanishes, leading to the ground truth becoming identical to the postulated prior.
By default, the simulation parameters are set as follows, unless otherwise specified:
$ T = 30 $,
$ N = 1000 $,
$ \alpha = 2 $,
$ c = 0.01 $,
$ v_{\Sf{T}} = v_{\Sf{F}} = 0.1 $,
$ \rho = 0 $.
For evaluating estimation accuracy, we use the MSE metric:
$
\Sf{MSE}
=
\frac{
	\|
		\hat{\bs{x}}
		-
		\bs{x}_{0}
	\|^{2}
}{
	\|
		\bs{x}_{0}
	\|^{2}
}
$.
We compare four algorithms in our study:
\begin{itemize}

\item
KVASP:
Detailed in Algo. \ref{Tab:VASP}, this algorithm uses the postulated distributions
$ q( x_{0} ) $
and
$ q( y | z_{0} ) $
as inputs;

\item
VAMP:
Described in \cite[Algo. 3]{rangan2019vector}, VAMP also takes the postulated distributions
$ q( x_{0} ) $
and
$ q( y | z_{0} ) $
as inputs;

\item
GASP:
Outlined in \cite[Algo. 1]{lucibello2019generalized}, GASP similarly uses the postulated distributions
$ q( x_{0} ) $
and
$ q( y | z_{0} ) $
as inputs;

\item
Bayes-optimal:
This is obtained by solving
$
\hat{\bs{x}}
=
\ArgMax\limits_{ \bs{x}_{0} }
p( \bs{x}_{0} )
p( \bs{y} | \bs{H} \bs{x}_{0} )
$.
This approach is generally NP-hard when the support of
$ p( \bs{x}_{0} ) $
is discrete \cite{rangan2012asymptotic}, making simulation infeasible in such cases.
However, when
$ p( \bs{x}_{0} ) $
is continuous, the Bayes-optimal estimator can be approximated using effective algorithms like VAMP, which utilize the ground-truth distributions
$ p( \bs{x}_{0} ) $
and
$ p( \bs{y} | \bs{z}_{0} ) $
as inputs \cite{takahashi2022macroscopic}.

\end{itemize}

\begin{figure}[!t]
\centering
\subfigure[i.i.d. matrix ($\rho=0$)]{
\label{Fig:VASP_1_iid}
\begin{minipage}[b]{.45\linewidth}
\centering
\includegraphics[scale=0.35]{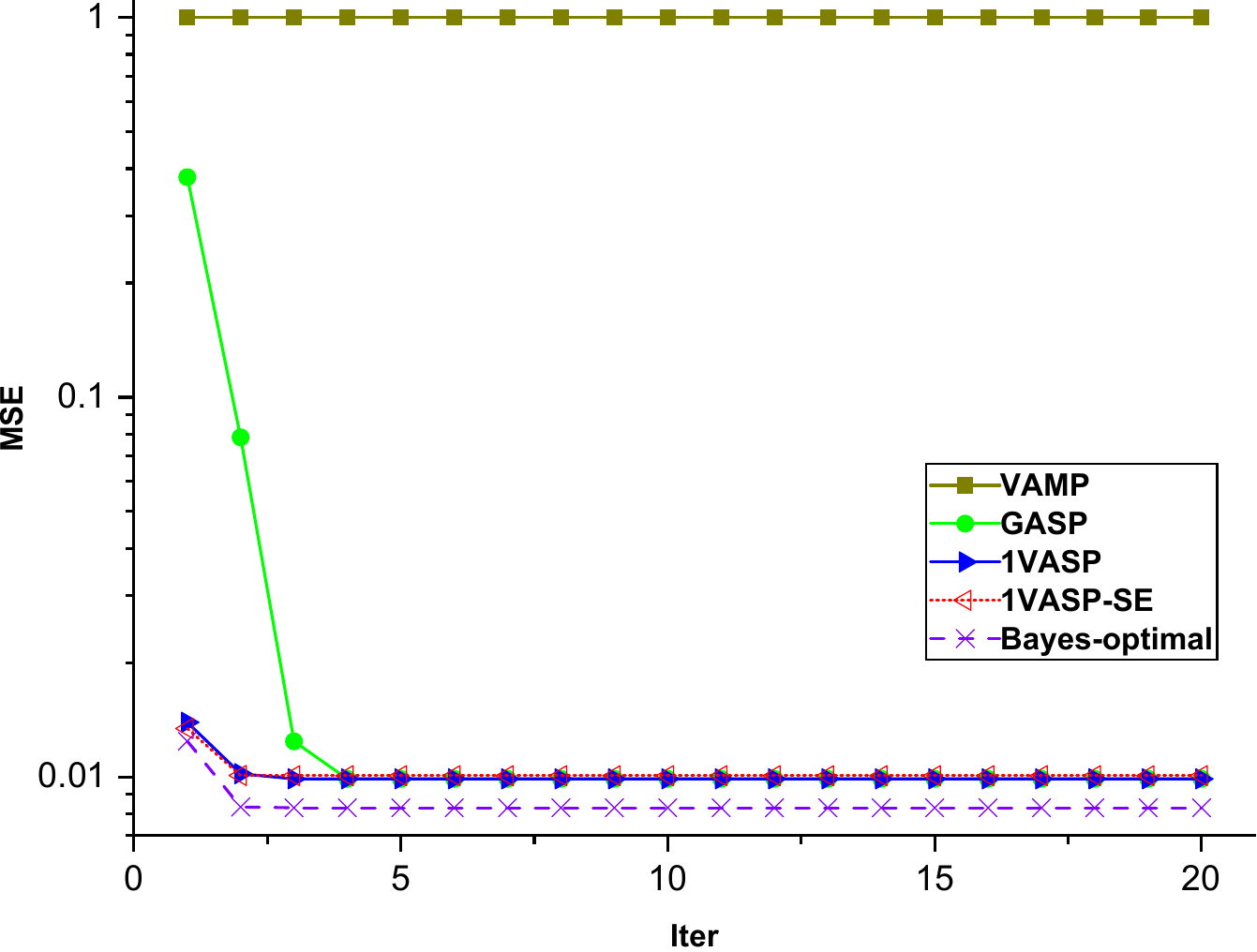}
\end{minipage}
}
\subfigure[correlated matrix ($\rho=0.4$)]{
\label{Fig:VASP_1_Correlated}
\begin{minipage}[b]{.45\linewidth}
\centering
\includegraphics[scale=0.35]{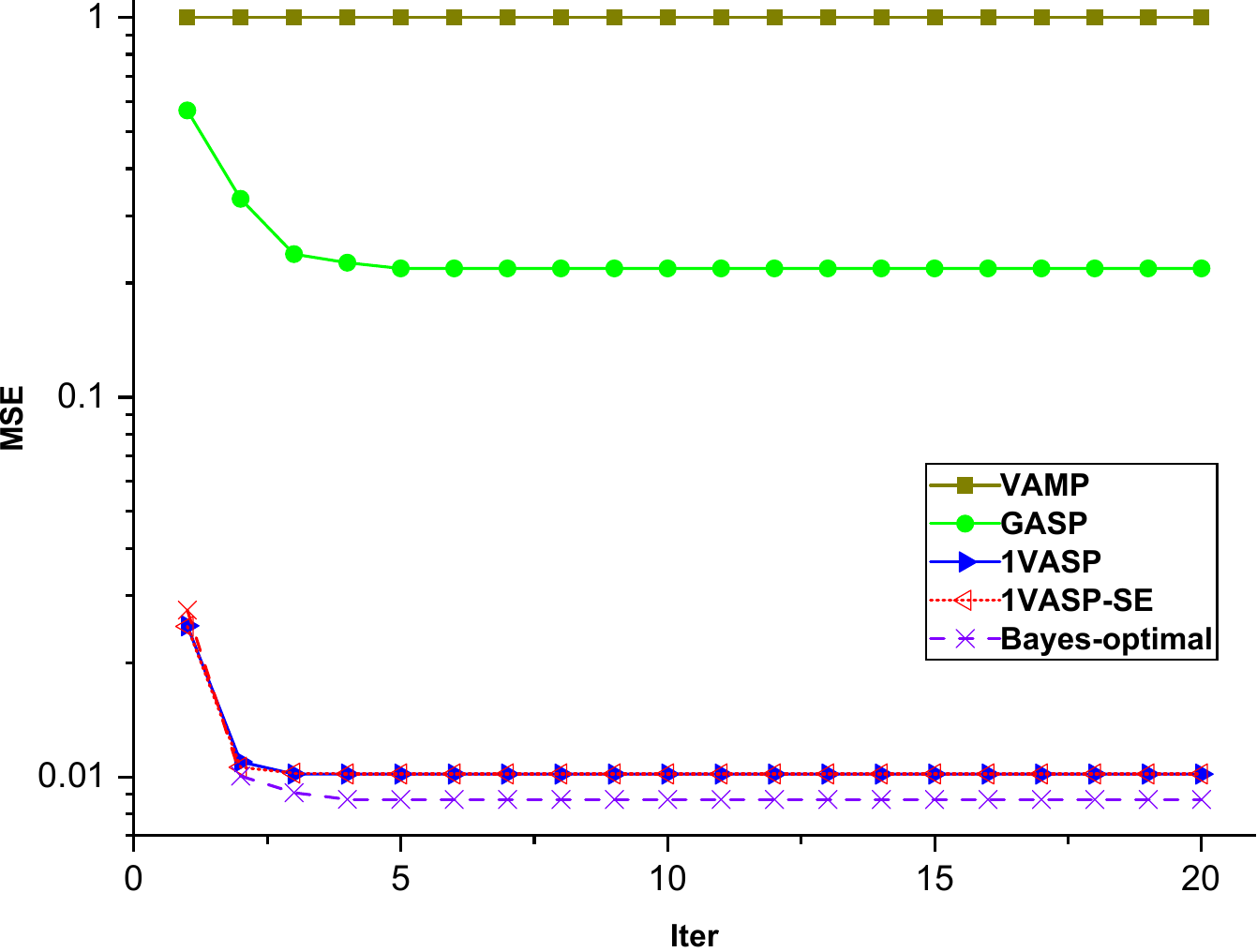}
\end{minipage}
}
\caption{
Comparison in model-mismatched cases ($ K = 1 $
and
$ L_{1} = 4 $)
}
\label{Fig:VASP_1}
\end{figure}

\begin{figure}[!t]
\centering
\subfigure[i.i.d. matrix ($\rho=0$)]{
\label{Fig:VASP_2_iid}
\begin{minipage}[b]{.45\linewidth}
\centering
\includegraphics[scale=0.3]{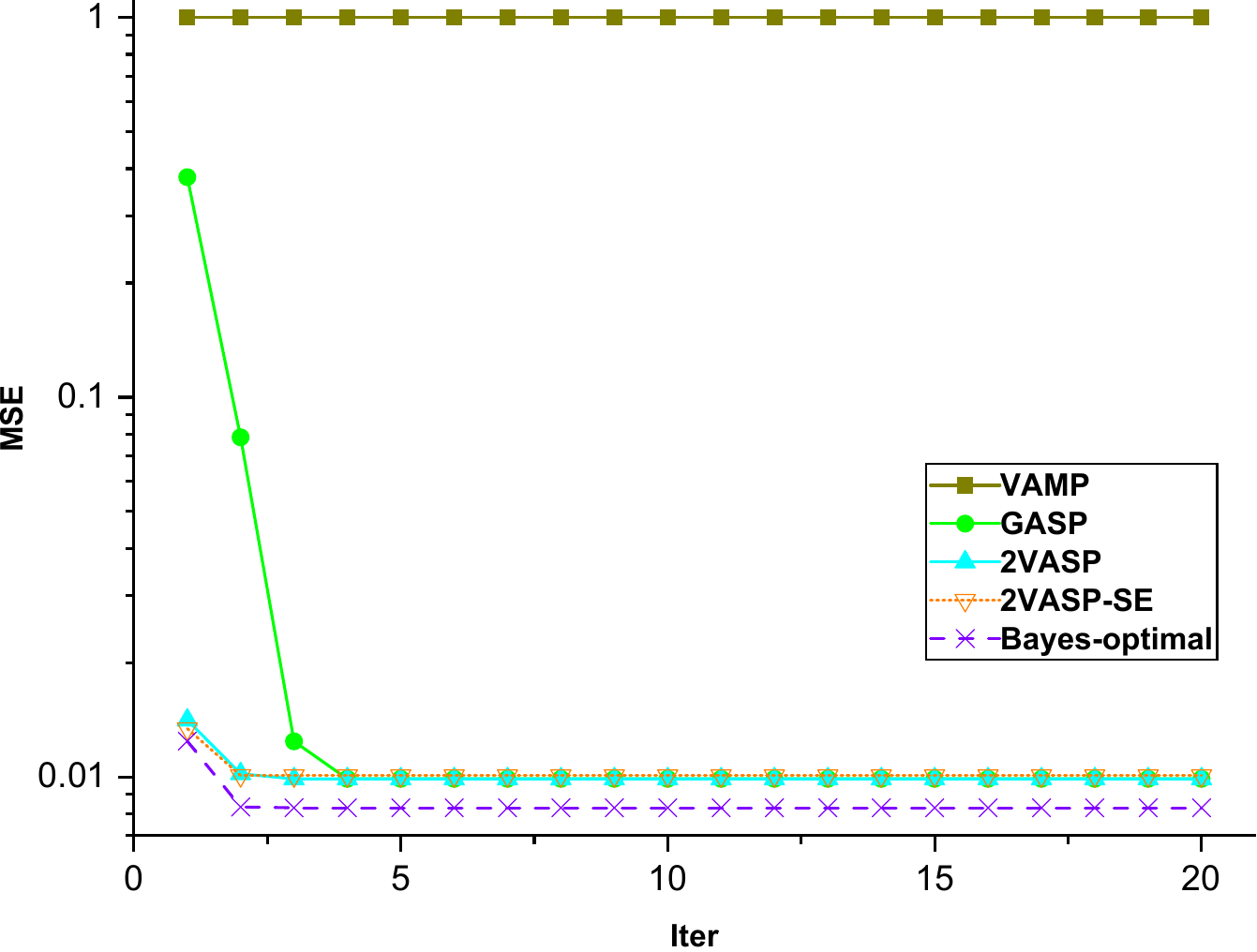}
\end{minipage}
}
\subfigure[correlated matrix ($\rho=0.4$)]{
\label{Fig:VASP_2_Correlated}
\begin{minipage}[b]{.45\linewidth}
\centering
\includegraphics[scale=0.3]{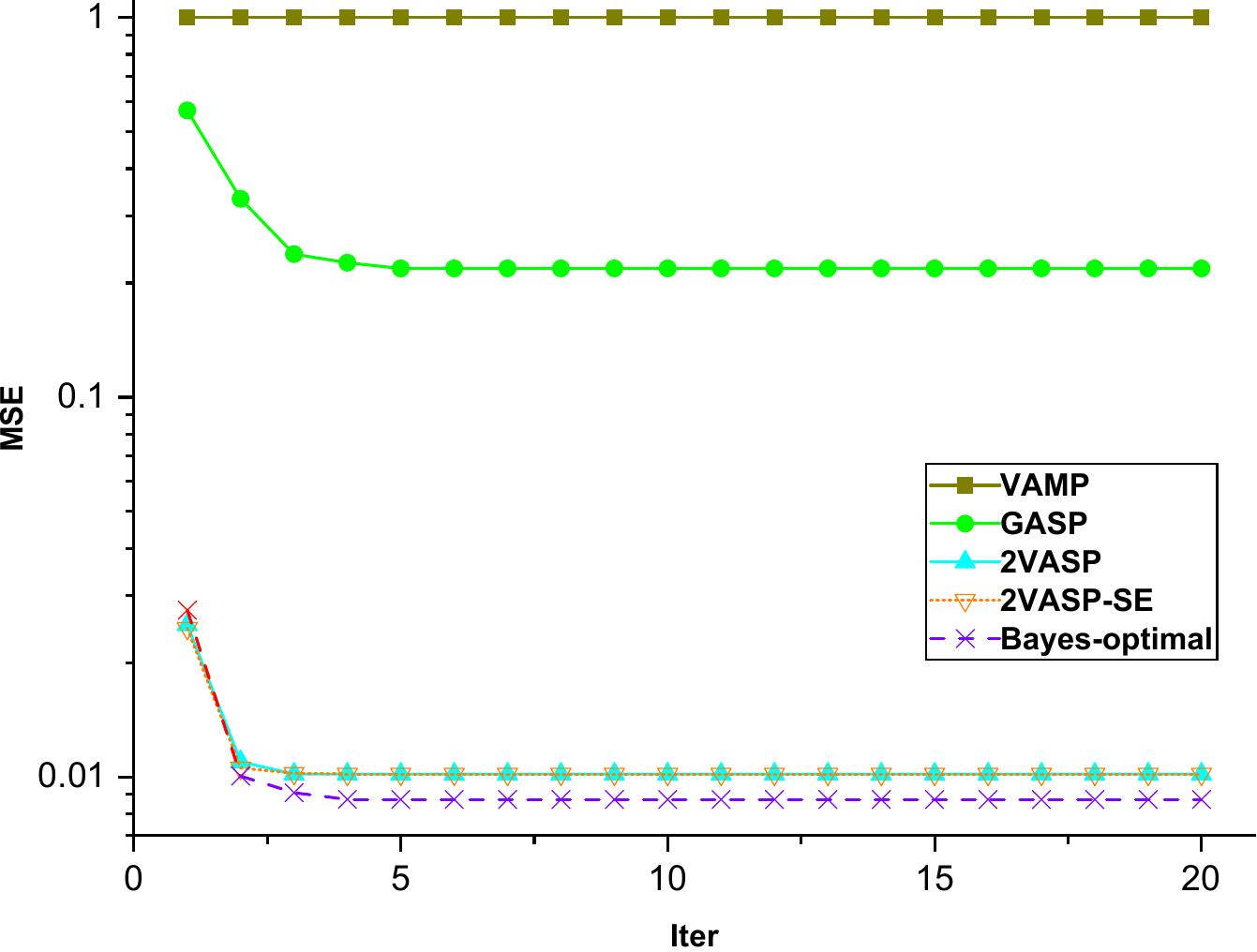}
\end{minipage}
}
\caption{
Comparison in model-mismatched cases ($ K = 2 $,
$ L_{1} = 2 $
and
$ L_{2} = 4 $)
}
\label{Fig:VASP_2}
\end{figure}

Fig. \ref{Fig:VASP_1} shows the results for
$ K = 1 $
and
$ L_{1} = 4 $, while
Fig. \ref{Fig:VASP_2} illustrates the results for
$ K = 2 $
,
$ L_{1} = 2 $
and
$ L_{2} = 4 $.
In the above cases, KVASP outperforms both VAMP and GASP, achieving an MSE that is close to that of the Bayes-optimal estimator.

\section{Conclusion}\label{sec:Conlcusion}

In the context of statistical inference under model mismatch, algorithms like VAMP \cite{takahashi2022macroscopic}, GASP \cite{lucibello2019generalized}, and VASP \cite{chen2023vector} can provide accurate estimates in certain scenarios.
However, many fundamental questions about model mismatch remain unresolved.
For instance, it is still unclear what degree of model mismatch is necessary for the PPE in \eqref{Eq:Postulated_Posterior_Estimator} to exhibit a RSB structure in the extremum conditions of its free energy \cite{takahashi2022macroscopic}, as well as what order of RSB is sufficient.
In this paper, we primarily contribute by developing a novel approximate message passing algorithm that incorporates KRSB.
This algorithm, referred to as KVASP, seamlessly reduces to VAMP and VASP under specific parameter choices.
Our work builds on impactful research in survey propagation related to sparse random constraint satisfaction problems \cite{mezard2002analytic,braunstein2005survey}.
The dynamics of KVASP are well characterized by the SE we present.
Notably, the SE fixed point equations align with the saddle points of the KRSB free energy.
This correspondence indicates that, once the KRSB ansatz is valid and the SE fixed point is attained, KVASP can accurately compute the PPE in \eqref{Eq:Postulated_Posterior_Estimator} in LSL.
This correspondence between SE and the KRSB analysis suggests that if the KRSB ansatz is correct and the algorithm converges to the SE fixed point, the proposed KVASP implements the PPE in \eqref{Eq:Postulated_Posterior_Estimator} with only cubic computational complexity.

\appendices

\section{Fixed Point Equations of the SE}\label{Appendix:Fixed_Point}

In the following, we analyze the fixed point equations of the SE.
To do this, we first remove the iteration indices
$ t $
and
$ t + 1 $
from Algo. \ref{Tab:SE}, since once a fixed point is reached, the results remain unchanged.
We then simplify the expressions as follows:
\begin{itemize}

\item
Introduce auxiliary parameters
$
\hat{\chi}_{\Sf{1x}}
\triangleq
\frac{1}{
	c_{\Sf{x, 0}}^{-}
}
$,
$
\hat{H}_{\Sf{1x, k}}
\triangleq
-
\frac{1}{ L_{k} }
(
	\frac{1}{
		c_{\Sf{x, k}}^{-}
	}
	-
	\frac{1}{
		c_{\Sf{x, k - 1}}^{-}
	}
)
$,
$
\hat{\chi}_{\Sf{2x}}
\triangleq
\frac{1}{
	c_{\Sf{x, 0}}^{+}
}
$,
$
\hat{H}_{\Sf{2x, k}}
\triangleq
-
\frac{1}{ L_{k} }
(
	\frac{1}{
		c_{\Sf{x, k}}^{+}
	}
	-
	\frac{1}{
		c_{\Sf{x, k - 1}}^{+}
	}
)
$,
$
\hat{\chi}_{\Sf{1z}}
\triangleq
\frac{1}{
	c_{\Sf{z, 0}}^{+}
}
$,
$
\hat{H}_{\Sf{1z, k}}
\triangleq
-
\frac{1}{ L_{k} }
(
	\frac{1}{
		c_{\Sf{z, k}}^{+}
	}
	-
	\frac{1}{
		c_{\Sf{z, k - 1}}^{+}
	}
)
$,
$
\hat{\chi}_{\Sf{2z}}
\triangleq
\frac{1}{
	c_{\Sf{z, 0}}^{-}
}
$,
$
\hat{H}_{\Sf{2z, k}}
\triangleq
-
\frac{1}{ L_{k} }
(
	\frac{1}{
		c_{\Sf{z, k}}^{-}
	}
	-
	\frac{1}{
		c_{\Sf{z, k - 1}}^{-}
	}
)
, \,
k \in [1, K]
$;

\item
Merge Lines 12-14 of Algo. \ref{Tab:SE} into Lines 35-37, and obtain
$
D_{\Sf{z}}^{-}
=
D_{\Sf{z}}^{+}
$,
$
F_{\Sf{z}}^{-}
=
F_{\Sf{z}}^{+}
$,
$
\hat{\mathtt{c}}_{\Sf{z, k}}^{-}
=
\hat{\mathtt{c}}_{\Sf{z, k}}^{+}
, \,
k \in [0, K]
$;

\item
Define
$
D_{\Sf{z}}
\triangleq
D_{\Sf{z}}^{-}
$,
$
F_{\Sf{z}}
\triangleq
F_{\Sf{z}}^{-}
$,
$
\chi_{\Sf{z}}
\triangleq
\hat{\mathtt{c}}_{\Sf{z, 0}}^{-}
$,
$
H_{\Sf{z, k}}
\triangleq
\frac{1}{ L_{k} }
(
	\hat{\mathtt{c}}_{\Sf{z, k}}^{-}
	-
	\hat{\mathtt{c}}_{\Sf{z, k - 1}}^{-}
)
, \,
k \in [1, K]
$;

\item
Simplify Lines 4-14 and 32-37 of Algo. \ref{Tab:SE} into Lines 12-16, 22-26, and 31-34 of Algo. \ref{Tab:Saddle_Point};

\item
Merge Lines 18-20 of Algo. \ref{Tab:SE} into Lines 29-31 of the same algorithm;

\item
define
$
D_{\Sf{x}}
$,
$
F_{\Sf{x}}
$,
$
\chi_{\Sf{x}}
$,
$
H_{\Sf{x, k}}
, \,
k \in [1, K]
$
similarly;

\item
Simplify Lines 15-31 of Algo. \ref{Tab:SE} into Lines 7-11, 17-21, and 27-30 of Algo. \ref{Tab:Saddle_Point}.

\end{itemize}

\section{Equations for MIMO Detection Problem}

By substituting
$q( x_{0} )$
into Line 19 of Tab. \ref{Tab:notations} in main text, we obtain the scalar output channel as:
$
f_{\Sf{x, 0}}(
	m_{\Sf{x, 0}},
	v_{\Sf{x, 0}}
)
=
-
\frac{
	(
		a - | m_{\Sf{x, 0}} |
	)^{2}
}{
	2 v_{\Sf{x, 0}}
}
$.
Next, by substituting
$
f_{\Sf{x, 0}}(
	m_{\Sf{x, 0}},
	v_{\Sf{x, 0}}
)
$
into Line 20, we can get the following equation as:
\begin{align*}
g_{\Sf{x, 1}}
& =
\frac{1}{2}
\log
v_{\Sf{x, 0}}
-
\frac{1}{2}
\log
(
	v_{\Sf{x, 0}}
	+
	L
	v_{\Sf{x, 1}}
)
+
\log
(
	Z_{ \Sf{+} }
	+
	Z_{ \Sf{-} }
)
,
\end{align*}
where
\begin{align*}
Z_{ \Sf{+} }
& \triangleq
\EXP[
	-
	\frac{
		L_{1}
		(
			m_{\Sf{x, 1}} - a
		)^{2}
	}{
		2
		(
			v_{\Sf{x, 0}}
			+
			L_{1} v_{\Sf{x, 1}}
		)
	}
]
\Phi[
	\frac{
		L_{1} v_{\Sf{x, 1}} a
		+
		v_{\Sf{x, 0}} \mu_{\Sf{x}}
	}{
		\sqrt{
			v_{\Sf{x, 1}}
			v_{\Sf{x, 0}}
			(
				v_{\Sf{x, 0}}
				+
				L_{1} v_{\Sf{x, 1}}
			)
		}
	}
]
, \\
Z_{ \Sf{-} }
& \triangleq
\EXP[
	-
	\frac{
		L_{1}
		(
			m_{\Sf{x, 1}} + a
		)^{2}
	}{
		2
		(
			v_{\Sf{x, 0}}
			+
			L_{1} v_{\Sf{x, 1}}
		)
	}
]
\Phi[
	\frac{
		L_{1} v_{\Sf{x, 1}} a
		-
		v_{\Sf{x, 0}} \mu_{\Sf{x}}
	}{
		\sqrt{
			v_{\Sf{x, 1}}
			v_{\Sf{x, 0}}
			(
				v_{\Sf{x, 0}}
				+
				L_{1} v_{\Sf{x, 1}}
			)
		}
	}
]
,
\end{align*}
and
$ \Phi[ \cdot ] $
is the cumulative distribution function of the standard normal distribution.

\section{A Useful Lemma}\label{Appendix:Summary}

\begin{lemma}\label{lemma:Decom_Formula}
A matrix
$
\bs{Q}
=
\mathcal{Q}(
	C, D, F, \chi,
	\{
		H_{k}
	\}_{k = 1}^{K}
)
\triangleq
F \mathbb{1}_{\tau}
+
\chi \Eye_{\tau}
+
\sum_{k = 1}^{K}{}
H_{k} \Eye_{
	\frac{ \tau }{
		\tilde{L}_{k}
	}
}
\otimes
\mathbb{1}_{
	\tilde{L}_{k}
}
$
in the KRSB form can be decomposed as described in \cite{shinzato2008perceptron}
\begin{align*}
\bs{Q}
& =
\bs{F} \bs{D}_{\Sf{q}} \bs{F}^{\Cts}
, \,
\bs{F}
\triangleq
\left[
\begin{array}{ cccc }
	1
	&
	\\
	&
	\frac{1}{
		\sqrt{
			\frac{\tau}{
				\tilde{L}_{K}
			}
		}
	}
	\bs{F}_{
		\frac{\tau}{
			\tilde{L}_{K}
		}
	}
	\otimes
	\frac{1}{
		\sqrt{
			\frac{
				\tilde{L}_{K}
			}{
				\tilde{L}_{K - 1}
			}
		}
	}
	\bs{F}_{
		\frac{
			\tilde{L}_{K}
		}{
			\tilde{L}_{K - 1}
		}
	}
	\otimes
	\cdots
	\otimes
	\frac{1}{
		\sqrt{
			\frac{
				\tilde{L}_{2}
			}{
				\tilde{L}_{1}
			}
		}
	}
	\bs{F}_{
		\frac{
			\tilde{L}_{2}
		}{
			\tilde{L}_{1}
		}
	}
	\otimes
	\frac{1}{
		\sqrt{
			\tilde{L}_{1}
		}
	}
	\bs{F}_{
		\tilde{L}_{1}
	}
	\\
\end{array}
\right]
, \\
\bs{D}_{ \Sf{q} }^{ (i, j) }
& \triangleq
\begin{cases}
	C
	&
	i = j = 0
	\\
	\sqrt{\tau} D
	&
	( i = 0, \, j = 1 )
	\,
	\text{or}
	\,
	( i = 1, \, j = 0 )
	\\
	\bs{D}_{\Sf{a}}^{
		(i - 1, i - 1)
	}
	&
	(
		i \geq 1, j \geq 1, i = j
	)
	\\
	0
	&
	\text{others}
	\\
\end{cases}
,
\end{align*}
where
$
\bs{D}_{\Sf{a}}
\triangleq
\Diag[
	\tau F \bs{e}_{\tau}
	+
	\chi \bs{1}_{\tau}
	+
	\sum_{k = 1}^{K}{}
	\tilde{L}_{k} H_{k}
	\bs{1}_{
		\frac{\tau}{
			\tilde{L}_{k}
		}
	}
	\otimes
	\bs{e}_{
		\tilde{L}_{k}
	}
]
$
and
$
\bs{F}_{L}
$
represents an
$
L \times L
$
discrete Fourier transform matrix.
\end{lemma}

\begin{table*}[!t]
\centering
\caption{Useful Notations}
\label{Tab:notations}
\scriptsize
\begin{tabular}{ l l l }
\toprule[1pt]
\TB{Line}
&
\TB{
Parameters associated with
$\bs{z}_{0}$
}
&
\TB{
Parameters associated with
$\bs{x}_{0}$
}
\\ \midrule[0.5pt]
\rownumber
&
$
v_{\Sf{z, 0}}
\triangleq
\frac{1}{
	\hat{\chi}_{\Sf{1z}}
}
$
&
$
v_{\Sf{x, 0}}
\triangleq
\frac{1}{
	\hat{\chi}_{\Sf{1x}}
}
$
\\
\rownumber
&
$
v_{\Sf{z, 1}}
\triangleq
\frac{1}{ L_{1} }
(
	\frac{1}{
		\hat{\chi}_{\Sf{1z}}
		-
		L_{1} \hat{H}_{\Sf{1z, 1}}
	}
	-
	\frac{1}{
		\hat{\chi}_{\Sf{1z}}
	}
)
$
&
$
v_{\Sf{x, 1}}
\triangleq
\frac{1}{ L_{1} }
(
	\frac{1}{
		\hat{\chi}_{\Sf{1x}}
		-
		L_{1} \hat{H}_{\Sf{1x, 1}}
	}
	-
	\frac{1}{
		\hat{\chi}_{\Sf{1x}}
	}
)
$
\\
\rownumber
&
$
v_{\Sf{z, k}}
\triangleq
\frac{1}{ L_{k} }
(
	\frac{1}{
		\hat{\chi}_{\Sf{1z}}
		-
		\sum_{i = 1}^{k}
		L_{i} \hat{H}_{\Sf{1z, i}}
	}
	-
	\frac{1}{
		\hat{\chi}_{\Sf{1z}}
		-
		\sum_{i = 1}^{k - 1}
		L_{i} \hat{H}_{\Sf{1z, i}}
	}
)
, \,
k \in [ 2, K ]
$
&
$
v_{\Sf{x, k}}
\triangleq
\frac{1}{ L_{k} }
(
	\frac{1}{
		\hat{\chi}_{\Sf{1x}}
		-
		\sum_{i = 1}^{k}
		L_{i} \hat{H}_{\Sf{1x, i}}
	}
	-
	\frac{1}{
		\hat{\chi}_{\Sf{1x}}
		-
		\sum_{i = 1}^{k - 1}
		L_{i} \hat{H}_{\Sf{1x, i}}
	}
)
$
\\
\rownumber
&
$
g_{\Sf{z, 0}}
\triangleq
\log
\int{\dd \tilde{z}} \,
q^{\beta}(y | \tilde{z})
\Normal[
	\tilde{z} | m_{\Sf{z, 0}},
	\frac{1}{\beta}
	v_{\Sf{z, 0}}
]
$
&
$
g_{\Sf{x, 0}}
\triangleq
\log
\int{
	\dd \tilde{x}
} \,
\Normal[
	\tilde{x}
	| m_{\Sf{x, 0}},
	\frac{1}{\beta}
	v_{\Sf{x, 0}}
]
q^{\beta}(
	\tilde{x}
)
$
\\
\rownumber
&
$
g_{\Sf{z, 1}}
\triangleq
\log
\int{
	\dd m_{\Sf{z, 0}}
} \,
\Normal[
	m_{\Sf{z, 0}} | m_{\Sf{z, 1}},
	v_{\Sf{z, 1}}
]
\EXP[
	\tilde{L}_{1}
	g_{\Sf{z, 0}}
]
$
&
$
g_{\Sf{x, 1}}
\triangleq
\log
\int{
	\dd m_{\Sf{x, 0}}
} \,
\Normal[
	m_{\Sf{x, 0}} | m_{\Sf{x, 1}},
	v_{\Sf{x, 1}}
]
\EXP[
	\tilde{L}_{1}
	g_{\Sf{x, 0}}
]
$
\\
\rownumber
&
$
g_{\Sf{z, k}}
\triangleq
\log
\int{
	\dd m_{\Sf{z, k - 1}}
} \,
\Normal[
	m_{\Sf{z, k - 1}} | m_{\Sf{z, k}},
	v_{\Sf{z, k}}
]
\EXP[
	\frac{
		L_{k}
	}{
		L_{k - 1}
	}
	g_{\Sf{z, k - 1}}
]
, \,
k \in [ 2, K - 1 ]
$
&
$
g_{\Sf{x, k}}
\triangleq
\log
\int{
	\dd m_{\Sf{x, k - 1}}
} \,
\Normal[
	m_{\Sf{x, k - 1}} | m_{\Sf{x, k}},
	v_{\Sf{x, k}}
]
\EXP[
	\frac{
		L_{k}
	}{
		L_{k - 1}
	}
	g_{\Sf{x, k - 1}}
]
$
\\
\rownumber
&
$
g_{\Sf{z, K}}
\triangleq
\log
\int{
	\dd m_{\Sf{z, K - 1}}
} \,
\Normal[
	m_{\Sf{z, K - 1}} | \mu_{\Sf{z}},
	v_{\Sf{z, K}}
]
\EXP[
	\frac{
		L_{K}
	}{
		L_{K - 1}
	}
	g_{\Sf{z, K - 1}}
]
$
&
$
g_{\Sf{x, K}}
\triangleq
\log
\int{
	\dd m_{\Sf{x, K - 1}}
} \,
\Normal[
	m_{\Sf{x, K - 1}} | \mu_{\Sf{x}},
	v_{\Sf{x, K}}
]
\EXP[
	\frac{
		L_{K}
	}{
		L_{K - 1}
	}
	g_{\Sf{x, K - 1}}
]
$
\\
\rownumber
&
$
\langle a \rangle_{\Sf{z, 0}}
\triangleq
\frac{
	\int{\dd \tilde{z}} \,
	q^{\beta}(y | \tilde{z})
	\Normal[
		\tilde{z} | m_{\Sf{z, 0}},
		\frac{1}{\beta}
		v_{\Sf{z, 0}}
	]
	a
}{
	\EXP[
		g_{\Sf{z, 0}}
	]
}
$
&
$
\langle a \rangle_{\Sf{x, 0}}
\triangleq
\frac{
	\int{
		\dd \tilde{x}
	} \,
	\Normal[
		\tilde{x}
		| m_{\Sf{x, 0}},
		\frac{1}{\beta}
		v_{\Sf{x, 0}}
	]
	q^{\beta}(
		\tilde{x}
	)
	a
}{
	\EXP[
		g_{\Sf{x, 0}}
	]
}
$
\\
\rownumber
&
$
\langle a \rangle_{\Sf{z, 1}}
\triangleq
\frac{
	\int{
		\dd m_{\Sf{z, 0}}
	} \,
	\Normal[
		m_{\Sf{z, 0}} | m_{\Sf{z, 1}},
		v_{\Sf{z, 1}}
	]
	\EXP[
		\tilde{L}_{1}
		g_{\Sf{z, 0}}
	]
	a
}{
	\EXP[
		g_{\Sf{z, 1}}
	]
}
$
&
$
\langle a \rangle_{\Sf{x, 1}}
\triangleq
\frac{
	\int{
		\dd m_{\Sf{x, 0}}
	} \,
	\Normal[
		m_{\Sf{x, 0}} | m_{\Sf{x, 1}},
		v_{\Sf{x, 1}}
	]
	\EXP[
		\tilde{L}_{1}
		g_{\Sf{x, 0}}
	]
	a
}{
	\EXP[
		g_{\Sf{x, 1}}
	]
}
$
\\
\rownumber
&
$
\langle a \rangle_{\Sf{z, k}}
\triangleq
\frac{
	\int{
		\dd m_{\Sf{z, k - 1}}
	} \,
	\Normal[
		m_{\Sf{z, k - 1}} | m_{\Sf{z, k}},
		v_{\Sf{z, k}}
	]
	\EXP[
		\frac{
			L_{k}
		}{
			L_{k - 1}
		}
		g_{\Sf{z, k - 1}}
	]
	a
}{
	\EXP[
		g_{\Sf{z, k}}
	]
}
, \,
k \in [ 2, K - 1 ]
$
&
$
\langle a \rangle_{\Sf{x, k}}
\triangleq
\frac{
	\int{
		\dd m_{\Sf{x, k - 1}}
	} \,
	\Normal[
		m_{\Sf{x, k - 1}} | m_{\Sf{x, k}},
		v_{\Sf{x, k}}
	]
	\EXP[
		\frac{
			L_{k}
		}{
			L_{k - 1}
		}
		g_{\Sf{x, k - 1}}
	]
	a
}{
	\EXP[
		g_{\Sf{x, k}}
	]
}
$
\\
\rownumber
&
$
\langle a \rangle_{\Sf{z, K}}
\triangleq
\frac{
	\int{
		\dd m_{\Sf{z, K - 1}}
	} \,
	\Normal[
		m_{\Sf{z, K - 1}} | \mu_{\Sf{z}},
		v_{\Sf{z, K}}
	]
	\EXP[
		\frac{
			L_{K}
		}{
			L_{K - 1}
		}
		g_{\Sf{z, K - 1}}
	]
	a
}{
	\EXP[
		g_{\Sf{z, K}}
	]
}
$
&
$
\langle a \rangle_{\Sf{x, K}}
\triangleq
\frac{
	\int{
		\dd m_{\Sf{x, K - 1}}
	} \,
	\Normal[
		m_{\Sf{x, K - 1}} | \mu_{\Sf{x}},
		v_{\Sf{x, K}}
	]
	\EXP[
		\frac{
			L_{K}
		}{
			L_{K - 1}
		}
		g_{\Sf{x, K - 1}}
	]
	a
}{
	\EXP[
		g_{\Sf{x, K}}
	]
}
$
\\
\multirow{2}*{\rownumber}
&
$
a_{\Sf{z}}
\triangleq
p(y | z_{0})
\Normal\left[
	z_{0}
	\left|
		\frac{
			\hat{D}_{\Sf{1z}}
			(
				\hat{\chi}_{\Sf{1z}}
				-
				\sum_{k = 1}^{K}{}
				L_{k} \hat{H}_{\Sf{1z, k}}
			)
		}{
			\hat{F}_{\Sf{1z}}
			(
				\hat{C}_{\Sf{1z}}
				+
				\frac{
					\hat{D}_{\Sf{1z}}^{2}
				}{
					\hat{F}_{\Sf{1z}}
				}
			)
		}
		\mu_{\Sf{z}},
		\frac{1}
		{
			\hat{C}_{\Sf{1z}}
			+
			\frac{
				\hat{D}_{\Sf{1z}}^{2}
			}{
				\hat{F}_{\Sf{1z}}
			}
		}
	\right.
\right]
\times
$
&
\multirow{2}*{
	$
	a_{\Sf{x}}
	\triangleq
	\Normal[
		\mu_{\Sf{x}} |
		\frac{
			\hat{D}_{\Sf{1x}}
		}{
			\hat{\chi}_{\Sf{1x}}
			-
			\sum_{k = 1}^{K}{}
			L_{k} \hat{H}_{\Sf{1x, k}}
		}
		x_{0},
		\frac{
			\hat{F}_{\Sf{1x}}
		}{
			(
				\hat{\chi}_{\Sf{1x}}
				-
				\sum_{k = 1}^{K}{}
				L_{k} \hat{H}_{\Sf{1x, k}}
			)^{2}
		}]
	p(x_{0})
	$
}
\\
&
$
\, \, \,
\Normal\left[
	\mu_{\Sf{z}}
	\left|
		0, \frac{
			\hat{F}_{\Sf{1z}}
			(
				\hat{C}_{\Sf{1z}}
				+
				\frac{
					\hat{D}_{\Sf{1z}}^{2}
				}{
					\hat{F}_{\Sf{1z}}
				}
			)
		}{
			\hat{C}_{\Sf{1z}} (
				\hat{\chi}_{\Sf{1z}}
				-
				\sum_{k = 1}^{K}{}
				L_{k} \hat{H}_{\Sf{1z, k}}
			)^{2}
		}
	\right.
\right]
$
&
\\ \midrule[0.5pt]
\rownumber
&
$
g_{\Sf{z}}
\triangleq
\frac{1}{ L_{K} }
g_{\Sf{z, K}}
$
&
$
g_{\Sf{x}}
\triangleq
\frac{1}{ L_{K} }
g_{\Sf{x, K}}
$
\\
\rownumber
&
$
\Gamma_{\Sf{z, 0}}
\triangleq
\frac{2}{L_{1} - 1}
(
	\frac{
		\partial g_{\Sf{z}}
	}{
		\partial v_{\Sf{z, 1}}
	}
	-
	L_{1}
	\frac{
		\partial g_{\Sf{z}}
	}{
		\partial v_{\Sf{z, 0}}
	}
)
$
&
$
\Gamma_{\Sf{x, 0}}
\triangleq
\frac{2}{L_{1} - 1}
(
	\frac{
		\partial g_{\Sf{x}}
	}{
		\partial v_{\Sf{x, 1}}
	}
	-
	L_{1}
	\frac{
		\partial g_{\Sf{x}}
	}{
		\partial v_{\Sf{x, 0}}
	}
)
$
\\
\rownumber
&
$
\Gamma_{\Sf{z, k}}
\triangleq
\frac{2}{
	\frac{
		L_{k + 1}
	}{
		L_{k}
	}
	- 1
}
(
	\frac{
		\partial g_{\Sf{z}}
	}{
		\partial v_{\Sf{z, k + 1}}
	}
	-
	\frac{
		L_{k + 1}
	}{
		L_{k}
	}
	\frac{
		\partial g_{\Sf{z}}
	}{
		\partial v_{\Sf{z, k}}
	}
)
, \,
k \in [ 1, K - 1 ]
$
&
$
\Gamma_{\Sf{x, k}}
\triangleq
\frac{2}{
	\frac{
		L_{k + 1}
	}{
		L_{k}
	}
	- 1
}
(
	\frac{
		\partial g_{\Sf{x}}
	}{
		\partial v_{\Sf{x, k + 1}}
	}
	-
	\frac{
		L_{k + 1}
	}{
		L_{k}
	}
	\frac{
		\partial g_{\Sf{x}}
	}{
		\partial v_{\Sf{x, k}}
	}
)
$
\\
\rownumber
&
$
\Gamma_{\Sf{z, K}}
\triangleq
-
2 \frac{
	\partial g_{\Sf{z}}
}{
	\partial v_{\Sf{z, K}}
}
+
L_{K} (
	\frac{
		\partial g_{\Sf{z}}
	}{
		\partial \mu_{\Sf{z}}
	}
)^{2}
$
&
$
\Gamma_{\Sf{x, K}}
\triangleq
-
2 \frac{
	\partial g_{\Sf{x}}
}{
	\partial v_{\Sf{x, K}}
}
+
L_{K} (
	\frac{
		\partial g_{\Sf{x}}
	}{
		\partial \mu_{\Sf{x}}
	}
)^{2}
$
\\
\rownumber
&
$
\hat{\mu}_{\Sf{z}}
\triangleq
c_{\Sf{z, K}}
\frac{
	\partial g_{\Sf{z}}
}{
	\partial \mu_{\Sf{z}}
}
+
\mu_{\Sf{z}}
$
&
$
\hat{\mu}_{\Sf{x}}
\triangleq
c_{\Sf{x, K}}
\frac{
	\partial g_{\Sf{x}}
}{
	\partial \mu_{\Sf{x}}
}
+
\mu_{\Sf{x}}
$
\\
\rownumber
&
$
\hat{c}_{\Sf{z, k}}
\triangleq
c_{\Sf{z, k}}
-
c_{\Sf{z, k}}^{2}
\Gamma_{\Sf{z, k}}
, \,
k \in [ 0, K ]
$
&
$
\hat{c}_{\Sf{x, k}}
\triangleq
c_{\Sf{x, k}}
-
c_{\Sf{x, k}}^{2}
\Gamma_{\Sf{x, k}}
$
\\ \midrule[0.5pt]
\rownumber
&
$
f_{\Sf{z, 0}}
\triangleq
\max\limits_{ \tilde{z} }
[
	\log q( y | \tilde{z} )
	-
	\frac{1}{
		2 v_{\Sf{z, 0}}
	}
	(
		\tilde{z} - m_{\Sf{z, 0}}
	)^{2}
]
$
&
$
f_{\Sf{x, 0}}
\triangleq
\max\limits_{ \tilde{x} }
[
	\log q( \tilde{x} )
	-
	\frac{1}{
		2 v_{\Sf{x, 0}}
	}
	(
		\tilde{x} - m_{\Sf{x, 0}}
	)^{2}
]
$
\\
\rownumber
&
$
g_{\Sf{z, 1}}
=
\log
\int{
	\dd m_{\Sf{z, 0}}
} \,
\Normal[
	m_{\Sf{z, 0}} | m_{\Sf{z, 1}},
	v_{\Sf{z, 1}}
]
\EXP[
	L_{1} f_{\Sf{z, 0}}
]
$
&
$
g_{\Sf{x, 1}}
=
\log
\int{
	\dd m_{\Sf{x, 0}}
} \,
\Normal[
	m_{\Sf{x, 0}} | m_{\Sf{x, 1}},
	v_{\Sf{x, 1}}
]
\EXP[
	L_{1} f_{\Sf{x, 0}}
]
$
\\
\rownumber
&
$
\Gamma_{\Sf{z, 0}}
=
- 2
(
	\frac{
		\partial g_{\Sf{z}}
	}{
		\partial v_{\Sf{z, 1}}
	}
	-
	L_{1}
	\frac{
		\partial g_{\Sf{z}}
	}{
		\partial v_{\Sf{z, 0}}
	}
)
$
&
$
\Gamma_{\Sf{x, 0}}
=
- 2
(
	\frac{
		\partial g_{\Sf{x}}
	}{
		\partial v_{\Sf{x, 1}}
	}
	-
	L_{1}
	\frac{
		\partial g_{\Sf{x}}
	}{
		\partial v_{\Sf{x, 0}}
	}
)
$
\\ \bottomrule[1pt]
\end{tabular}
\end{table*}

\ifCLASSOPTIONcaptionsoff
	\newpage
\fi

\bibliographystyle{IEEEtran}
\normalem
\bibliography{IEEEabrv,./CQ_Original,./CQ_bib}

\end{document}